\documentclass[preprint,aps,showpacs]{revtex4}

\usepackage{graphicx}
\usepackage{dcolumn}
\usepackage{bm}
\usepackage{amsfonts}
\usepackage{amsfonts}

\begin{document}


\title{Stability and mode analysis of solar  coronal loops  using
thermodynamic irreversible energy principles}
\author{A. Costa}
\email{ acosta@mail.oac.uncor.edu; costa@iafe.uba.ar}
\affiliation{Instituto de Astronom\'\i a y F\'\i sica del Espacio
(CONICET-Argentina) }
\author{Rafael Gonz\'{a}lez }
\affiliation{ Universidad
Nacional de General Sarmiento (UNGS) and Departamento de F\'\i
sica (FCEyN-UBA-Argentina) }
\date{\today}

\begin{abstract}
We study the modes and stability of non - isothermal
coronal loop models with different intensity values of the
equilibrium magnetic field. We use an energy principle obtained
via non - equilibrium thermodynamic arguments. The principle is
expressed in terms of Hermitian operators and allow to consider
together the coupled system of equations: the balance of energy
equation and the equation of motion. We determine modes
characterized as long - wavelength disturbances that are present
in inhomogeneous media. This character of the system introduces
additional difficulties for the stability analysis because the
inhomogeneous nature of the medium determines the structure of the
disturbance, which is no longer sinusoidal. Moreover, another
complication is that we obtain a continuous spectrum of stable
modes in addition to the discrete one. We obtain a unique
unstable mode with a characteristic time that is comparable with
the characteristic life--time observed for loops. The feasibility
of wave--based and flow--based models is examined. 

\end{abstract}
\pacs{Valid PACS appear here}
\maketitle


\section{\label{sec:intro}Introduction}
\subsection{\label{sec:level2}Variational principles}

Stability is a crucial requirement for a model to produce
realistic descriptions. Thus, different stability analyzed of
solar structures can be found in the literature, generally
restricted to special types of perturbations and specific
equilibrium models. These includes, models that consider adiabatic
configuration such as the ones analyzed via the classical
criterion of Bernstein et al. \cite{bers} or those that
presuppose  static equilibrium and analyze thermal stability. In
the application of Bernstein's
 criterion, the adiabatic assumption implies that the energy balance equation is not required
 and thus dissipation is
impossible. Also the assumption of static models is a strong, and
often unjustified, restriction for open systems. Thus, a crucial
question for any theoretical model is whether  the much more
common far--from--equilibrium states are stable, where the
consideration of both thermal and mechanical coupled equations
must be included.

A more realistic analysis of the  stability of configurations
represented by non-conservative equations was presented by Lerche
and Low \cite{ler2}. They proposed a Lagrangian principle in
order to analyze quiescent prominences that can undergo thermal
instabilities. However the non--self--adjoint character of the
operators involved in the obtained principle makes   the physical
interpretation difficult.

 In this paper we apply an energy
principle to analyze the stability of solar coronal loops. The
principle was obtained in a previous paper (Paper I: Costa et al.
\cite{us00}; see also Sicardi et al. \cite{us0}, see also
Sicardi et al. \cite{us, sica, sic, sicar}) using a general
procedure of irreversible thermodynamics -based on firmly
established thermodynamic laws- that can be understood as an
extension of  Bernstein's MHD principle to situations far from
thermodynamic equilibrium. This fact has the advantage that many
known results obtained by simpler criteria can be re--examined by
a direct comparison with our criterion, and that, as it is
obtained via a thermodynamic approach, allows a straightforward
physical interpretation. The principle associates stability with
the sign of a quadratic form avoiding non--self--adjoint
operators. Obtaining a self--adjoint operator is a requirement for
our principle to hold. When this is accomplished the calculus is
simplified. The self--adjoint character of an operator implies
that the eigenvalues $\omega^{2}$ are  real. Hence stability
transitions always occur when $\omega^{2}$ crosses zero, rather
than at  particular points of the real axis where the real part of
the eigenvalue is different from zero, i.e.  $Re ( \omega) \neq
0$, leading to an efficient formulation to test stability. Thus,
the symmetry considerations of the self--adjoint operators, the
fact that there is a diagonal form associated with these
operators, and that the Rayleigh--Ritz theorem states the
existence of a minimum eigenvalue, are important reasons to try to
maintain self--adjointness in the consideration of stability.

\subsubsection{\label{sec:level3}Solar coronal loops}

The theoretical modeling and the interpretation of observations of
coronal loop systems deal, among others, with the  discussion
 whether the propagating observed disturbances in loops and
post--flare loops are waves or plasma flow.

Dynamical features of brightening coronal loops have been
traditionally interpreted as field--aligned flow of matter
generated by  asymmetries in the energy input. Most classical
theoretical models have difficulties in determining the physical
conditions that make them compatible with observations. Both
static loops and steady state models -for the two classes of
temperatures models: hot (isothermal coronas with T $\approx
10^{6}K$) and cool (gradually increasing temperatures up to T
$\approx 10^{5}K$) - fail to provide a satisfactory explanation
for both the emission measure distribution and the Doppler shift
observations  (Jordan \cite{jor}; Serio \cite{ser}; Craig
and McClymont \cite{cra}; Mariska \cite{mar}). Thus, this
suggests that in
  traditional model scenarios  radiative losses cannot
be compensated by thermal conduction.  Therefore, other heating
mechanisms must be assumed (Aschwanden et al. \cite{asch1};
\cite{asch2}; Walsh and Galtier \cite{walsh}). Also,
theoretical time--dependent models of individual loops where the
plasma evolves in response to a cyclical process of heating and
cooling of the flow have difficulties in fitting observations
(Klimchuk and Mariska \cite{klim}).

The assumption of propagating disturbances associated with slow
magnetoacoustic waves in high Alfv\'en speed media is also a field
of investigation. Several wave--based models were developed to
explain  observations (Nakariakov et al. \cite{naka1};
Tsiklauri and Nakariakov \cite{naka2}).  These authors suggest
that  -depending on the relative importance of dissipation by
magnetic resistivity- upwardly propagating waves  (of observed
periods between 5-20 min) that decay significantly in the vicinity
of the loop apex could explain  the rarity of observational
detection of downwardly propagating waves. However, upwardly
propagating disturbances with non-decaying or even growing
amplitudes were observed in coronal EIT plumes. Analytic models
have shown
 that slow magnetoacoustic waves may be trapped and
nonlinearly steepened with height, providing a possible
interpretation of the phenomenon (Ofman et al \cite{naka3}).

However, due to the intensity of the flaring, the plasma dynamic
of flare loops is generally  associated with flows rather than
with waves. In fact, systematic intensity perturbations in
post--flare loops can suggest that they are the result of
evaporation--condensation cycles caused by the efficient heating
of the flaring plasma from the chromosphere.
 Thus, chromospheric evaporation seems to be the main initial
matter inflow source for flare loops. De Groof et al. \cite{deg}
analyzed an off--limb half loop structure from an EIT shutterless
campaign  and gave arguments to reject the slow magnetoacoustic
description and to support the flowing/falling plasma one.
Nevertheless, these authors admit that the wave theory cannot be
excluded yet.

Other authors have suggested that a combination of phenomena can
be at the basis of a better interpretation. Alexander et al.
\citep{ale} examined 10 flares and  concluded that plasma
turbulence could be the source of the observed intensity changes
rather than hydrodynamic flows such as chromospheric evaporation.
They pointed out that it cannot be excluded that there is a degree
of "gentle evaporation" occurring early in the event with
associated hard X--ray emission below their threshold of
detection. A series of more recent papers (Tsiklauri et al.
\cite{tsi1}a; \cite{tsi2}b; \cite{tsi3}c) that combine
theoretical and observational analysis showed that oscillations in
white, radio and X--ray light curves observed during solar and
stellar flares may be produced by slow standing
 magnetoacoustic modes of the loops. They found that a
transient heat deposition at the loop bottom -or at the apex-
leads to a posterior up--flow evaporation of material of the order
of a few hundreds of $km / s^{-1}$. During the peak of the flare,
the simulations showed that a combined action of heat input and
conductive and radiate losses could yield an oscillatory pattern
with typical amplitudes of up to a few tens of $km / s^{-1}$.
Then, a cooling phase of plasma draining  with velocities of the
order of hundreds of $km / s^{-1}$ occurs. The numerical
quasi--periodic oscillations in all the physical quantities, that
resemble observational features, were interpreted  as being
produced by standing sound waves caused by impulsive and localized
heating.

In previous papers (Borgazzi and Costa \cite{cos1}; Costa and
Stenborg \cite{cos2}) one of us developed a diagnostic
observational method to describe loop intensity variations, both
in space and time, along coarse--grain loop structures. We find
that none of the arguments leading to the determination of whether
waves or flow models can better fit observations was conclusive.
Some of our results suggested wave--based model interpretations
i.e. the periodic behaviour of the disturbances observed, the
almost constant speed of some brightening features and the fact
that the estimated speeds were not higher that the sound speed in
the coronal loops. However, as we mentioned, the period behaviour
can also be attributed to flows (G\'omez et al. \citep{gom1}; De
Groof et al. \citep{deg}). Also, even when the calculated speeds
were not greater than the sound ones,
 some of the velocity patterns were far from being constant
and their values were comparable to the free-fall case.

Another open question is the relation between the loop's coronal
dynamics and the physical conditions on the chromospheric bases.
Borgazzi and Costa \citep{cos1} found a longitude of chromospheric
coherence that characterizes the behaviour of a whole loop--system
of evolving coronal--isolated filaments. This description is in
accordance with limit--cycle models that require that the
triggering mechanism of the dynamics is located at the bottom of
the structure giving rise to the observed similar coronal
conditions of the isolated filaments. Another aspect that deserves
attention is whether it is physically possible that the
periodicity observed could be related to, or could be the
consequence of propagating magnetoacoustic modes from the
chromosphere that have suffered distortion due to the dispersing
media.

Other point that is under debate is the thermal structure of the
loops. Loop observations with TRACE (Transition Region and Coronal
Explorer, Handy et al. \citep{Handy}) suggest that hot coronal
loops are isothermal and more dense than the predictions of static
loop models. However this
 scenario is not conclusive and other
interpretations are possible.  Reale and Peres \citep{rea} showed
that bundles of thin strands, each one behaving as a static loop,
with its characteristic thermal structure, convoluted with the
TRACE temperature response could appear as a single almost
isothermal loop. A wide range of configurations can be proposed to
fit observations. The fact that images form a compound of complex
integrated time--varying data that are not easy to resolve is at
the basis of this difficulty. The loops under analysis are
surrounded by other structures that usually intersect them along
the line of sight and the change of the brightening of the loops
is also affected by  background emission. Thus,  efforts are made
to produce observational and theoretical results  of coronal loop
dynamics.

The aim of this paper is to investigate  whether the propagating
observed disturbances in loops are waves or plasma flow and their
thermal structure. Non--isothermal loops are traditional
candidates for Hopf instabilities with cycles of flow evaporating
and condensing, thus the analysis of frequencies and mode
structures can provide insight into a possible wave model
interpretation of these types of configurations. We consider the
stability analysis as the leading criterion to select possible
theoretical wave models. The fact that a number of non--linear
equilibria are possible due to the open character of the systems
 makes it necessary to consider both thermal and
mechanical stability in a coupled way.

\section{The stability criterion}

\noindent The thermodynamics of irreversible processes is
described in terms of phenomenological relations between conjugate
pairs of thermodynamic variables: the flows and the forces that
cause them. The linear thermodynamic approximation treats small
deviations from the equilibrium state by including fluctuations in
the neighborhood of this state. It describes the behaviour of the
system around the equilibrium state or around a non--equilibrium
stationary one that is linearly close to it.

If the system is isolated, as is stated by the second law of
thermodynamics, the entropy grows exponentially up to its maximum
value. That the system is in an open--near--equilibrium state
means that energy and matter is exchanged with the neighbors and
the entropy of the system is not necessarily positive. Even when
the entropy produced in the system's interior, due to irreversible
processes, is never negative, a negative flow of entropy produced
by the exchange of matter and energy can make the system  remain
indefinitely in a near--equilibrium state. These states are known
as stationary states and a coherent dynamic of the system could
last if sufficient negative entropy flow is provided to it. Thus,
the criterion that states the stability of this stationary state
gives insight into the dynamic structures that can be found in
nature. These stationary states are also known as detailed
balanced. As Onsager pointed out \citep{ons1}, the balance
consists of the compensation between the fluctuations and
dissipation produced by the flows and forces that have a
microscopic reversible character near the thermodynamic
equilibrium. The empirical relations between flows and forces are
linear and related by the so--called resistance matrix
$\mathbf{R}$ that is symmetric and positive definite. Its
symmetric character is guaranteed by the principle of microscopic
reversibility and its positive definiteness by the proximity of
the reference state to the thermodynamic equilibrium, where the
entropy has a maximum.

However, there is no continuity between linear and nonlinear
thermodynamical processes. When the system is beyond the immediate
neighborhood of the stationary state the nonlinearities become
visible. Instabilities that cause dynamic transitions in open
systems are responsible for the qualitative difference between
linear and nonlinear thermodynamics. Therefore, dynamic
cooperative phenomena can only arise in nonlinear thermodynamics.
Thus, nonlinear thermodynamics is related to the stability
properties of non--equilibrium stationary states, where the linear
relation between flows and forces can become state dependent (i.e.
$\mathbf{R}$ is not necessarily a symmetric positive definite
matrix), and the problem of having a thermodynamic theory to
provide a general criterion for the stability of the system -which
is not evident through the integration of the variational
equations- becomes a fundamental point. Non--linear thermodynamics
is the extension of the linear theory to situations far from
thermodynamic equilibrium where the relaxation of the processes to
a steady state of non-equilibrium (nonlinear state) is not assured
and requires a stability analysis (Glansdorff and Prigogine
\cite{gl2}; Keizer \cite{kei}; Graham \cite{gra}; Lavenda
\cite{lav,laven}).

In Paper I we showed how to obtain the variational principle from
the equations that describe the dynamics of the system of
interest. The method consists of obtaining a Lyapunov function,
also known as generalized potential, that represents the
mathematical expression of the stability conditions. This function
is determined by the analysis of the thermodynamic properties of
the system linearized around a non--linear stationary state also
called non--linear equilibrium state. The equations governing the
dynamics are written as a system of two coupled equations: the
balance energy equation and the equation of motion. Thus, the
perturbation analysis around a stationary state is performed
considering a variable state vector of four independent
components: the three space component displacement and the
temperature variation. Once the linearization is done, the
Lyapunov function can be immediately obtained by inspection of the
resulting expression written in a compact matrix form. Each of the
matrices of the compact expression are linear operators (that
could include spatial derivatives) and have a clear physical
interpretation that is given by its role in the equation. The
matrix that multiplies the second time derivative of the
perturbation is associated with the inertia of the system, the one
that multiplies the first time derivative of the perturbation is
associated with dissipation and the one that multiplies the
perturbation itself is associated with potential forces over the
system. The principle is subject to physically reasonable
requirements of hermiticity and antihermiticity over the matrices.
For a more detailed presentation see Paper I and the references
presented there.

\subsection{The magnetohydrodynamic expression}

\noindent The specific model we analyze is taken to be composed of
a magnetohydrodynamic ideal plasma (i.e. with infinite electrical
conductivity $\sigma \gg 1 $). The fundamental ideal
magnetohydrodynamic equations to be considered are as follows. The
mass conservation equation,

\begin{equation}
\frac{\partial \rho}{\partial
t}+\nabla\cdot(\rho\vec{v})=0\label{1}
\end{equation}
where $\rho$ is the density of the plasma, $\vec{v}$ is the plasma
velocity, and $t$ the time. The perfect gas law or state equation,

\begin{equation}
 p=\frac{k_{B}}{m}\rho T \label{2}
\end{equation}
$k_{B}$ is the Boltzmann constant, $p$ the pressure, $T$ the
temperature and $m\equiv m_{p}$ the proton mass. For a fully
ionized $H$ plasma $\rho = \mu n_{e} m_{p}$; the solar coronal
abundances ($H:He=10:1$) correspond to a  molecular weight
$\mu=1.27$; $n_{e}$ is the number density of electron particles
(Aschwanden \citep{asch}. The induction equation,

\begin{equation}
\frac{\partial \vec{B}}{\partial
t}=\nabla\times(\vec{v}\times\vec{B})\label{3}
\end{equation}
$\vec{B}$ is the magnetic field vector. The magnetic diffusivity
was discarded. The equation of motion for the problem is:

\begin{equation}
\rho\frac{D\vec{v}}{Dt}=-\nabla(p)+\frac{1}{4\pi\mu}(\nabla\times\vec{B})\times\vec{B}
-\rho\nabla\phi\label{4}
\end{equation}
where $g=-\nabla\phi$ is the gravity expression and $j=1/4\pi
\nabla\times\vec{B}$ the current density. The energy balance
equation takes the form:
\begin{equation}
\frac{\rho^{\gamma}}{(\gamma-1)}\frac{D}{D
t}(\frac{p}{\rho^{\gamma}})=-L\label{5}
\end{equation}
where $ \gamma$ is the ratio of specific heats and $L$ is the
energy loss function:

\begin{equation}
  L=-\nabla\cdot\vec{F_{c}}-L_{r}+H.\label{6}
\end{equation}
$\vec{F_{c}}$ is the heat flux due to particle conduction along
the loop, $L_{r}$ is the net radiation flux. Neither the dominant
heating mechanism of coronal loops nor the spatial distribution
function of the energy input is known. So, the heating function is
the least known term in the energy balance equation. Thus, the
usual situation is to try reasonable arbitrary mathematical
functions which must fit the constraint imposed by the equilibrium
conditions.  As our model considers inhomogeneous temperature
gradients and isothermality is usually associated with footpoint
heated loops more than with uniformly heated ones we discard the
first case and tried the general expression $H = h \rho + H_{0}$.
A time varying dependence of $H$ was not considered for
simplicity. However, it could be a requirement for modeling
special events such as micro--flares or while considering magnetic
reconnection phenomena.
 Eq.~\ref{5} expresses the fact that the gain in particle
energy (internal plus kinetic) is due to heating sources, heat
flow and radiation losses; ohmic dissipation $j^{2}/\sigma$ and
all other heating sources were considered as vanishing terms
implying that the optically thin assumption holds. Then
$L_{r}=n_{e}n_{H}Q(T)$; the temperature variation ($Q(T)=\chi
T^{\alpha}$) was taken from Priest \citep{prie}. Also $F_{c}=-k
\nabla T$ and, as conduction across the magnetic field has been
discarded,  for a total ionized plasma
$F_{c}=-k_{0}T^{\frac{5}{2}}\nabla_{\parallel}T$. Finally equation
~\ref{5} can be written as

\begin{equation}
\frac{\rho^{\gamma}}{\gamma-1}\frac{D(\frac{p}{\rho^{\gamma}})}{Dt}=
\nabla\cdot(k_{0}T^{\frac{5}{2}}
\nabla_{\parallel}T)-\frac{\rho^{2}}{m^{2}} \chi
T^{\alpha}+\frac{\upsilon}{m^{2}}\rho  + H_{0}\label{7}
\end{equation}
where $\upsilon$ is a constant value to be determined from the
equilibrium conditions.

The linearization procedure is performed by replacing
$\rho=\rho_{0}+\rho_{1}$, $T=T_{0}+T_{1}$, $B=B_{0}+B_{1}$ and
$\vec{v}=\vec{v_{0}}+\partial\vec{\xi}/\partial t$ in the last
equations, and assuming hydrostatic conditions for the equation of
motion. Thus, $\vec{v_{0}}=0$ and
$\vec{v_{1}}=\partial\vec{\xi}/\partial t$ where $\xi$ is the
perturbation around the equilibrium of the equation of motion (the
stationary state), also $\partial\rho_{0}/\partial t=0$ and
$\partial B_{0}/\partial t=0$. Using the relation
$\partial/\partial t\simeq i\omega$ in eq.~\ref{1} and
eq.~\ref{3}, the corresponding linearized equations (eq.~\ref{8}
-~\ref{12}) are:

\begin{equation}
\rho_{1}+\nabla\cdot(\rho_{0}\vec{\xi})=0\label{8}
\end{equation}
\bigskip
\begin{equation}
p_{1}=\frac{k_{B}}{m}(\rho_{0}T_{1}-T_{0}\nabla\cdot(\rho_{0}\vec{\xi}))\label{9}
\end{equation}
\bigskip
\begin{equation}
\vec{B_{1}}=-\nabla\times(\vec{B}_{0}\times\vec{\xi})\label{10}
\end{equation}
\bigskip
\noindent
$\rho_{0}\ddot{\vec{\xi}}=\frac{k_{B}}{m}\nabla(T_{0}\nabla\cdot(\rho_{0}\vec{\xi})
-\rho_{0}T_{1})- $
\begin{equation}
- \frac{1}{4\mu}[(\nabla \times Q) \times B_{0}+ ( \nabla \times
B_{0} ) \times Q ] + \bigtriangledown \phi \nabla \cdot ( \rho_{0}
\vec{\xi })\label{11}
\end{equation}
\noindent or equivalently \noindent
$\rho_{0}\ddot{\vec{\xi}}-F\xi+\frac{k_{B}}{m}\nabla(\rho_{0}T_{1})=0$
\bigskip
\noindent and
\begin{equation}
 \frac{k_{B}}{m(\gamma-1)}[\rho_{0}\dot{T_{1}}-(\gamma-1)T_{0}
\nabla\cdot(\rho_{0}\dot{\vec{\xi}})]-AT_{1}+B\vec{\xi}=0
\label{12}
\end{equation}
\noindent being $$ A=-[c\nabla \cdot (T_{0}^{\frac{5}{2}}\nabla
_{\parallel }(\star)+\frac{5}{2}T_{0}^{\frac{3}{2}}\nabla
_{\parallel }(T_{0}))-\frac{\rho _{0}^{2}}{m^{2}}\chi \alpha
T_{0}^{\alpha -1}]  $$ \\ \noindent and $$
 B=\ \{\frac{k_{B}}{m}\beta \nabla _{\bullet }(\rho _{0}\
;\star )\} ; \ \beta =\ \frac{-2(\rho _{0}\chi T_{0}^{\alpha }-
\upsilon /2)}{k_{B}m}$$ \\ \noindent $$
 c=\frac{1.8 10^{-10}}{ln\Lambda} \ W m^{-1}K^{-1},  \;    \
Q=\vec{B}_{0}\times\vec{\xi}.$$ \\ The term $\ \nabla \cdot (\rho
_{0}\dot{\xi}) \ $ was discarded because it represents the total
net flux of material through the magnetic tube. The two obtained
equations  are expressed in terms of the displacement and
temperature perturbed variables $\vec{\xi}$ and $T_{1}$. $\star$
represents the location of the perturbed variables when performing
the matrix product.

Following Paper I the resulting energy principle is:

\bigskip
\noindent $$\delta^{2} S  = \frac{ 1}{ 2} \ \int [
\-\dot{\vec{\xi^{*}}}\beta \rho_{0}\dot{\vec{\xi}} d^{3}x \ + \ $$
\begin{equation}
+ \int  (  \vec{\xi}^{*} \beta  F \vec{\xi} +T_{1}^{*} AT_{1}
+T_{1}^{*}B\vec{\xi} -\vec{\xi}^{*}B T_{1})] d^{3}x \geq 0
\label{13}
\end{equation}
where F is the known Bernstein operator for the system.

For the non-dissipative cases, last expression reduces to the
well--known Bernstein MHD energy principle
\begin{equation}
\delta^{2} S  = \frac{ 1}{ 2}\int [\-\dot{\vec{\xi^{*}}}\beta
\rho_{0}\dot{\vec{\xi}} d^{3}x + \int  { \vec{\xi^{*}} \beta F
\vec{\xi}} ] d^{3}x \geq 0 \label{14}
\end{equation}
from where the eigenmodes and eigenfrequencies are calculated as
\begin{equation}
\omega^{2}=-\frac{\int \vec{\xi}^{*} \beta F\vec{\xi}
d^{3}x}{\int\vec{\xi}^{*}\beta \rho_{0}\vec{\xi} d^{3}x }
 \label{15}
\end{equation}
and  the stability criterion is obtained by requiring the
positivity of the potential energy of the perturbation (Galindo
Trejo \citep{gal1})
\begin{equation}
 \delta^{2} W_{p(Bernstein)} = \frac{1}{2}\int \vec{\xi}^{*} \beta F\vec{\xi} d^{3}x
 \label{16}
\end{equation}
subject to the normalization condition that the total kinetic
energy is equal to one. Thus, the dissipative principle and the
new frequencies are respectively:
\begin{equation}
 \delta^{2} W_{p} =\frac{1}{ 2}\int ( \vec{\xi}^{*} \beta  F \vec{\xi}+T_{1}^{*} AT_{1}
 +T_{1}^{*}B\vec{\xi} -\vec{\xi}^{*}BT_{1})d^{3}x\geq 0.
 \label{17}
\end{equation}
\begin{equation}
\omega^{2}  =- \frac{\int  ( \vec{\xi}^{*}  \beta  F\vec{\xi} +
T_{1}^{*}AT_{1}+T_{1}^{*}B\vec{\xi}-\vec{\xi}^{*}BT_{1}^{*} )
d^{3}x}{\int (\vec{\xi^{*}} \beta \rho_{0}\vec{\xi} )d^{3}x}
\label{18}
\end{equation}
with the same normalization condition.

\section{Application to the stability of a coronal inhomogeneous loop model}

We are interested in analyzing the stability of non--homogeneous
loops. This is, loops with inhomogeneous distributions of plasma
density and temperatures. This character of the system poses
additional difficulties for the stability analysis because the
inhomogeneous nature of the medium determines the structure  of
the disturbance which is no longer sinusoidal, making the
traditional normal mode analysis useless for this treatment.
Moreover, there may exist a continuous spectrum of stable modes
besides the discrete one. As a first order approximation we
neglect the effect of gravitational stratification and thus
confine  the analysis to characteristic spatial scales lower than
the pressure scale height in the solar corona. In order to analyze
the stability and to obtain the frequencies and modes the physical
quantities in eq.~\ref{17} and eq.~\ref{18} must be calculated
along the loop structure.

\subsection{Mechanical equilibrium}

\begin{figure}[t]
 \includegraphics[width=4.cm]{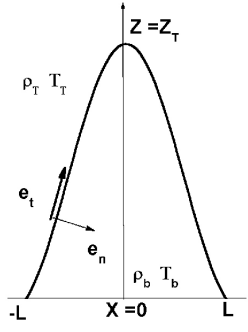}
   \caption{Schematic figure of the magnetic arcade with $z(x)= z_{t}+\frac{2L}{\pi}\ln\left[\cos(\pi
\frac{x}{2L})\right]$, $x$ and $z$ the Cartesian coordinates.
$e_{t}$ and $e_{n}$ the tangential and normal versors
respectively. $T_{b}$ and $T_{t}$ are the temperature values
  at the bottom and top  respectively. The same notation is used
 for the density $\rho$.}
   \label{fig:uno}
   \end{figure}

In order to determine an equilibrium configuration we assume
force--free equations due to the fact that in plasma with low
$\mathcal{\beta}$ (gas pressure over the magnetic pressure) the
pressure gradient can be neglected in comparison to the Lorentz
force. The coronal arcade is obtain from the equations

\begin{equation}
\nabla \times \mathbf{B_{0}}=\alpha \mathbf{B_{0}}=0 \label{19}
\end{equation}

\begin{equation}
\mathbf{j} \times \mathbf{B_{0}}=0. \label{20}
\end{equation}
Also, $\mathbf{B_{0}} \cdot \nabla p =0$ and thus the pressure has
a constant value along the loop. We assume that the unperturbed
magnetic field is $\mathbf{B}_{0} = (B_{0,x}(x,z), 0 ,
B_{0,z}(x,z))$ and obtain the equilibrium field components

\begin{equation}
B_{0x}= - B_{00} \cos(\frac{\pi}{2L} x) e^{-\frac{\pi}{2L}z}
\label{21}
\end{equation}
\begin{equation}
B_{0z}= B_{00} \sin(\frac{\pi}{2L} x) e^{-\frac{\pi}{2L}z}
\label{22}
\end{equation}
with
\begin{equation}\mathbf{B}_{0}= B_{00} e^{-\frac{\pi}{2L}z}\mathbf{e}_{s}. \label{23}
\end{equation}
The relation

\begin{equation}
z= z_{t}+\frac{2L}{\pi}\ln\left[\cos(\pi
\frac{x}{2L})\right]\label{24}
\end{equation}
is straightforward. The arc element $s$ (see Figure~\ref{fig:uno})
can be expressed as
\begin{equation}
ds= dx \sqrt{\left(1+\frac{dz}{dx}\right)^{2}} =  dx\; \triangle
\label{25}
\end{equation}
 with $\Delta= \sqrt{1+(z')^{2}}.$

\subsection{Thermal equilibrium}

The thermal equilibrium is obtained from  eq.~\ref{7} with $L=0$
(in eq.~\ref{6}). Thus expressions
$F_{c}=-k_{0}T^{\frac{5}{2}}\nabla_{\parallel}T$ satisfies the two
relations

\begin{equation}
\frac{\partial F_{c}}{\partial T} \frac{\partial T}{\partial s} =
\frac{-F_{c}}{k_{0}T^{\frac{5}{2}}} \frac{\partial F_{c}}{\partial
T }= -\frac{\rho^{2}}{m^{2}}Q(T)+ H, \label{26}
\end{equation}
from where we obtain the two equations $$F_{c}=\frac{- dT
k_{0}T^{\frac{5}{2}}}{ds} $$
\begin{equation}
\frac{{F_{c}}^{2}}{2}= {\int_{T_{0}}}^{T}
k_{0}T'^{\frac{5}{2}}\left[\frac{\rho^{2}}{m^{2}}Q(T')-H_{0}\right]
dT' \label{27}
\end{equation}
where we assume $F_{c}(s=0)=0$ as $dT/ds=0$  at $s=0$ and $H =
H_{0}$, so the constant value of  eq.~\ref{7} is $\upsilon =0$. We
then can replace $F_{c}$ in eqs.~\ref{27} and give $Q(T)$ its
explicit expression. Then integrating between $T_{t}$ and $T_{b}$
(the temperatures at the top and the bottom of the loop
respectively) and using $(dT/ds)_{T=T_{b}}=0$ and $T_{t}\gg T_{b}$
we obtain the constant value of the  heating function $H_{0}= 7
p^{2} \chi T^{\alpha -2}_{t}/(8 k_{B}^{2}\left(\alpha
+\frac{3}{2})\right)$. Also, we find

\begin{equation}
\left[\frac{dT}{ds}\right]^{2}=\frac{p^{2}\chi}{2k_{B}^{2}k_{0}(\alpha
+\frac{3}{2})}T^{\alpha-\frac{7}{2}} \left[1 - (
\frac{T}{T_{t}})^{2-\alpha}\right]\label{28}
\end{equation}
which is equivalent to the calculus in chapter 6 of Priest
\citep{prie}. Our aim is to obtain $T$ as a function of the line
element $s$. From eq.~\ref{28}   $s=f(T)$ given as an integral
expression of the temperature, which has to be inverted. Thus, for
calculus purposes, we define $v=1-(T/T_{t})^{2-\alpha}$ and
replace $T$ as a function of $v$ in eq.~\ref{28}. Then we obtain
$s=f(v)$ as
\begin{equation}
s=\frac{1}{\mathcal{A}}\mathbb{B}_{v}(\frac{1}{2},q)\label{29}
\end{equation}
 where 
\noindent
\begin{equation}
\mathbb{B}_{v}(\frac{1}{2},q)=\int_{0}^{v}t^{p-1}(1-t)^{q-1}dt
\nonumber \end{equation} \\ 

\noindent (Arfken and Weber \citealp{arf})
 with $$ p=\frac{1}{2}, \  \
q=(\frac{\alpha}{2}+\frac{3}{4})(2-\alpha)+1,$$ \\
$$\mathcal{A}=(2-\alpha)T_{t}^{\frac{\alpha}{2}-\frac{11}{4}}((p^{2}\chi)/(2k_{0}
(\alpha+\frac{3}{2})k_{B}^{2}))^{\frac{1}{2}}.$$  \\ Then,
$T=f^{-1}(s)$ as
\begin{equation}
\frac{dT}{ds}=\mathcal{A}\left[\frac{d\mathbb{B}_{v}}{dv}\frac{dv}{dT}\right]^{-1}\label{30}
\end{equation}

\subsection{The perturbation}

In order to calculate the stability and structure modes the
general perturbation expression along the equilibrium loop is
written
\begin{equation}
\vec{\xi}=[\zeta(x)
\mathbf{e}_{t}+\eta(x)\mathbf{e}_{n}+\xi_{y}(x)\mathbf{e}_{y}]e^{iky}\label{31}
\end{equation}
and
\begin{equation}
T_{1}=T_{1}(x)e^{iky}.\label{32}
\end{equation}

Then, representing the equilibrium functions of the different
quantities with a 0 sub-index and using the loop parameters and
the mathematical relations presented in the Appendix, we obtain
the non--dimensional expression for the energy principle
(eq.~\ref{17})

 $$ \delta^{2} W_{p}=\frac{1}{2}\int_{-1}^{1}dx
\left \{ \left [ \beta \frac{dT_{0}}{dx}f \left (
(\frac{d\rho_{0}}{dx})f+\rho_{0}D_{x}f-k\rho_{0}\xi_{y} \right )+
\right.  \right.$$ $$
 +\beta T_{0}f \left ( \frac{d^{2}\rho_{0}}{dx^{2}}f+ \rho_{0}
Dxxf- k \frac{d\rho_{0}}{dx}\xi_{y}-k \rho_{0}
\frac{d\xi_{y}}{dx}\right ) -$$
 $$-k \beta T_{0} \xi_{y}\left (
\frac{d\rho_{0}}{dx} f-k\rho_{0}\xi_{y} \right ) \left. \right.
{\Bigg ]} +
 $$
 $$
+C_{1}\left [ \beta \frac{d}{dx}  \left (
\frac{z'}{\triangle}\mathbf{B_{0}} \right ) \left (
\frac{k\mathbf{B_{0}}\xi_{y}}{\triangle} \left (
\frac{-z'}{\triangle}\zeta+\frac{\eta}{\triangle} \right )+
\right. \right.$$
 $$ + \left ( k\frac{z'}{\triangle}
\mathbf{B_{0}}\xi_{y}-\frac{d\mathbf{B_{0}}}{dx}\eta-\mathbf{B_{0}}\frac{d\eta}{dx}
\right ) \left ( \frac{\zeta}{\triangle}+ \eta\frac{z}{\triangle}
\right ) \left. \right )-\beta \left(\left (
k\frac{\mathbf{B_{0}}}{\triangle} \right )^{2}\xi_{y}^{2}+ \right.
$$
 $$ + \left ( \frac{d}{dx}(\frac{\mathbf{B_{0}}}{\triangle})
\xi_{y}+\frac{\mathbf{B_{0}}}{\triangle}\frac{d\xi_{y}}{dx} \right
)^{2}+( \frac{d\mathbf{B_{0}}}{dx}\eta+\mathbf{B_{0}}
\frac{d\eta}{dx}-k\frac{z'}{\triangle}\mathbf{B_{0}}\xi_{y})^{2}
{\Bigg )}  {\Bigg ]}+  $$
 $$+ C_{2}\left [ -\zeta T_{0}^{\frac{3}{2}}
\frac{1}{\triangle^{2}}\frac{dT_{0}}{dx}
T_{1}\frac{dT_{1}}{dx}-2\frac{T_{0}^{\frac{5}{2}}}{\triangle}\frac{d}{dx}
\left ( \frac{1}{\triangle} \right ) T_{1}\frac{dT_{1}}{dx} +
\right.
 $$ $$ -\frac{T_{0}^{\frac{5}{2}}}{\triangle^{2}} T_{1}
\frac{d^{2} T_{1}} {dx_{2}}
-
\frac{d}{dx} \left (
\frac{5}{2}T_{0}^{\frac{3}{2}}\frac{1}{\triangle^{2}}\frac{dT_{0}}{dx}
\right ) T_{1}^{2}  {\Bigg ]} -
\rho_{0}T_{0}^{\alpha-1}T_{1}^{2}+\beta\frac{d\rho_{0}}{dx}T_{1}f+$$
$$ +\beta\rho_{0} \left ( D_{x}f-k\xi_{y} \right ) T_{1}+$$

 \begin{equation}
-\beta \left [ \frac{1}{\triangle}\frac{d\rho_{0}}{dx}\zeta
T_{1}+\frac{\rho_{0}}{\triangle}\zeta\frac{dT_{1}}{dx}
+\frac{z'}{\triangle}\frac{d\rho_{0}}{dx}\eta\frac{dT_{1}}{dx}+k\rho_{0}\xi_{y}
T_{1} \right ]  {\Bigg \}} \label{33} 
\end{equation}
\noindent 
where the
non--dimensional quantity  $\delta^{2} W_{p}/\left(\chi
T_{t}^{\alpha+1}\rho_{t}^{2}L/m^{2}\right)$ replaces $\delta^{2}
W_{p}$, and $C_{1}=B_{00}^{2}k_{B} T_{t}\rho_{t}/m\mu$ and
$C_{2}=c \; m^{2} T_{t}^{\frac{7}{2}}/(L^{2}
T_{t}\alpha\rho_{t}^{2})$ were used. From this variational
principle we can then analyze stability and obtain the mode
structure and the associated frequencies for the general  mode
given by  eq. \ref{31}.

\section{Results and discussion}

In order to calculate modes and frequencies we followed the
schematic procedure described in Paper I and in Galindo Trejo
\citep{gal1}. We used a symbolic manipulation program to integrate
the equations. $\delta^{2}W_{p}$ and the perturbations were
expanded in a three dimensional--Fourier basis that adjusts to
border conditions. Thus, a quadratic form for $\delta^{2}W_{p}$
was obtained and was minimized with the Ritz variational
procedure. A matrix discrete eigenvalue problem subject to a
normalization constraint was obtained. The procedure is equivalent
to solving eq.~\ref{18} of our modified principle. Once the modes
are obtained, the stability condition of eq.~\ref{17} for the
generalized potential energy: $\delta^{2}W_{p}\geq0$ must be
corroborated. The following values were used for the numerical
calculation of the modes

$$\alpha=-\frac{1}{2} \ \ \rightarrow \ \ q=\frac{6}{5}$$ $$ s=
\frac{1}{\mathcal{A}}  \mathbb{B}_{v}(\frac{1}{2}, \frac{6}{5}) \
\ \rightarrow \ \mathcal{A}=\frac{5}{2}T_{t}^{3}(
\frac{p^{2}\chi}{2k_{0}k_{B}^{2}})^{ \frac{1}{2}}.$$

\noindent Coronal loop parameters: $L=10^{10}cm$ (or $L=100Mm$),
$T_{b}=10^{4}K$ $T_{t}=10^{6}K$ $n_{e}=10^{8}cm^{-3}$ electron
number density $p_{t} =2k_{B}T_{t} \;$;
$\rho_{t}=mp_{t}/k_{B}T_{t}$.

\bigskip

 Our main  concern was to know whether the
magnetic configuration of equilibrium could be  stable under
linear perturbations. For non homogeneous configurations it is
well known that the stable eigenvalues can have  continuous
spectra while the unstable ones have  a discrete spectrum (see
Freidberg \citep{fre} or Priest \citep{prie}). If the
resulting  mode components have a characteristic wavelength of the
order of the equilibrium structure, the non--homogeneous character
of modes could determine, for the stable modes, a continuous
spectrum.  Thus, in this case, the traditional normal mode
analysis gives only a rough description because one of the
consequence of the existence of the continuum is that an
accumulation of discrete eigenvalues can take place at either
boundary, generally at $\omega^{2}=0$ or $\omega^{2}=\infty$,
indicating the presence of a continuum stable spectrum. Note that
as  the  basis used is discrete, the spectrum is necessarily
discrete. However, the additional evaluation of the generalized
potential energy provides the correct unstable modes and gives
 an approximate value of the most probable stable period when the smaller $\omega^{2}$
 is not located at the boundaries.

We used different values for $k$: $k=0$, $k=0.5$ and $k=10$ ($k$
is the wavenumber associated with the perturbation component
transverse to the plane that contains  the magnetic
configuration). We also calculated the frequencies and modes for
two different values of the magnetic field: $B_{00}=11G$ and
$B_{00}=100G$.

Table 1 and Table 2 show the  eigenvalues (periods) associated
with the different modes for the cases $B_{00}=11G$ and
$B_{00}=100G$ respectively, considering $k=0$ and obtained by
solving eq.~\ref{18}. We obtained 12 eigenfrequencies and 12
eigenmodes for each of the magnetic field values i.e. we used a
three--component expansion and a four--component perturbation
vector. We evaluated the mode behaviour for $k\neq0$. For each
mode corresponding to a complex eigenvalue, the perturbation
$\xi_{y}$ was at least two orders of magnitude smaller that the
parallel $\zeta$ component and the normal $\eta$ components. For
the  modes with real eigenvalue, in only one case was $\xi_{y}$
comparable to the smaller of the two other spatial perturbations.
Thus, for numerical simplicity, we used $k=0$ and we discarded
three zero eigenvalues associated with this choice of $k$. We then
analyzed only  the 9 relevant modes. This means that a
two--dimensional analysis of the dynamics of the problem is
reasonably able to obtain the overall behaviour within the
approximations we are considering. Thus, we decided to investigate
the unstable modes and to consider the most stable one as a
reference value for stability. The most stable mode is the one
that has a real $\omega$ value and minimizes $\delta^{2}W_{p}$ (it
is the mode that gives the minimum positive value of the
functional $\delta^{2}W_{p}$) and the most unstable one is the
mode that  has a complex $\omega$ value with the minimum value of
$\tau \simeq 1/|\omega|$ ($\tau$ the instability time).

From the analysis of the table data we can conclude: 1) for each
of the two investigated magnetic values  we have three complex
values of  $\omega$  and six real ones;  2) in the two magnetic
field cases  the eigenvalues of the first mode are the same; 3) in
all the other cases the eigenvalues  with  $B_{00}=100G$ are
almost an order of magnitude smaller than the corresponding values
of $B_{00}=11G$;  4) the series of  eigenvalues is such that it
could be possible  that the stable periods accumulate at
$\omega=0$, thus the definite stability characterization is
subject to the evaluation of the generalized potential energy of
the
 modes.

 \begin{figure*}[t]
 \includegraphics[width=4.4cm]{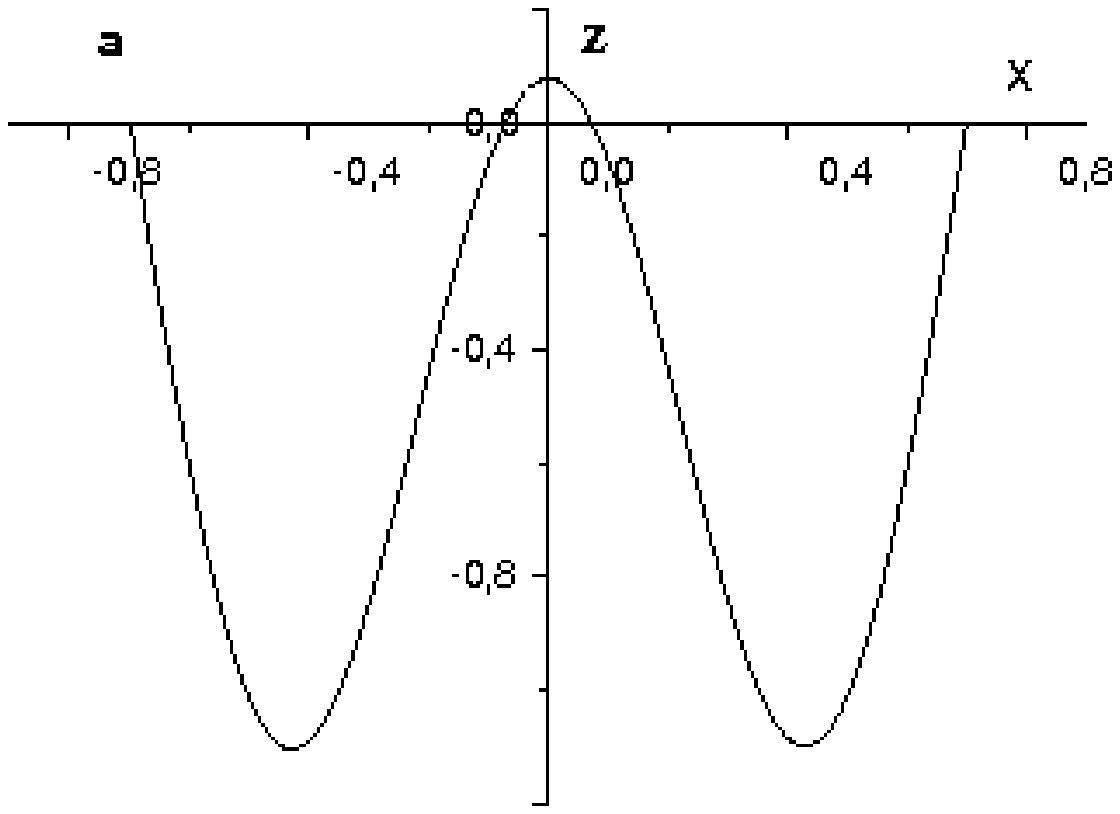}
 \includegraphics[width=4.4cm]{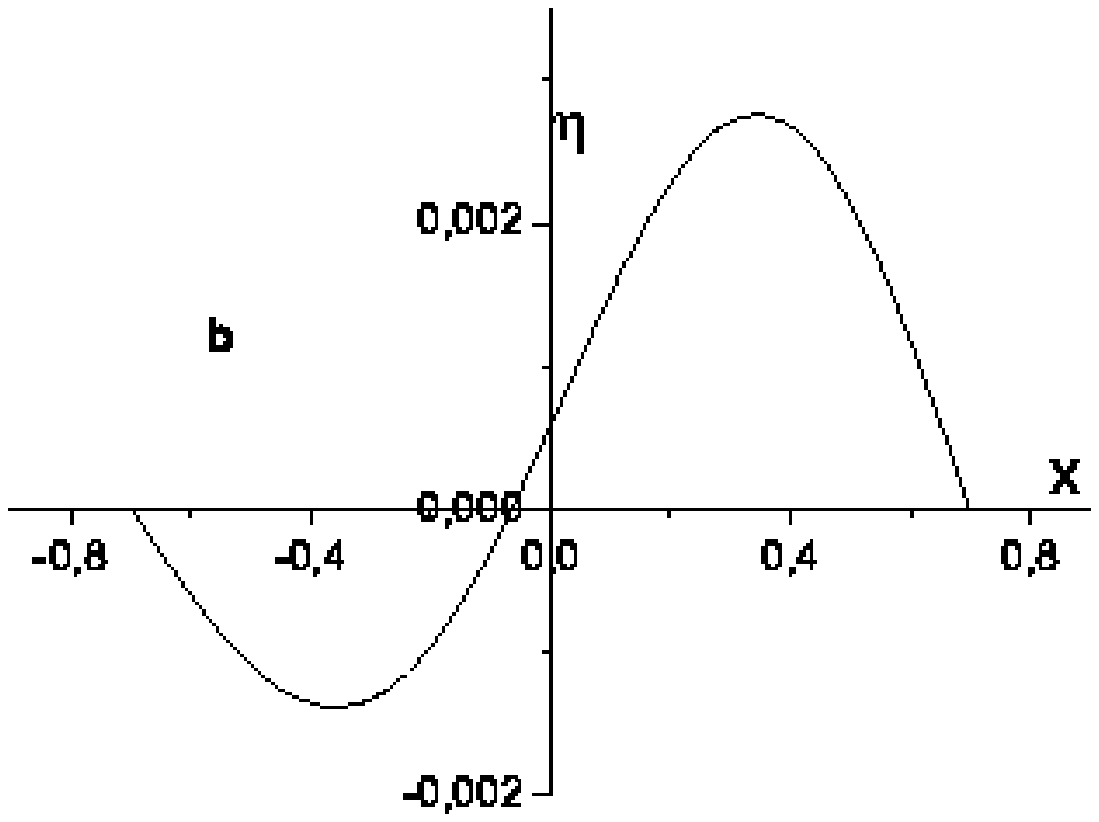}
  \includegraphics[width=4.4cm]{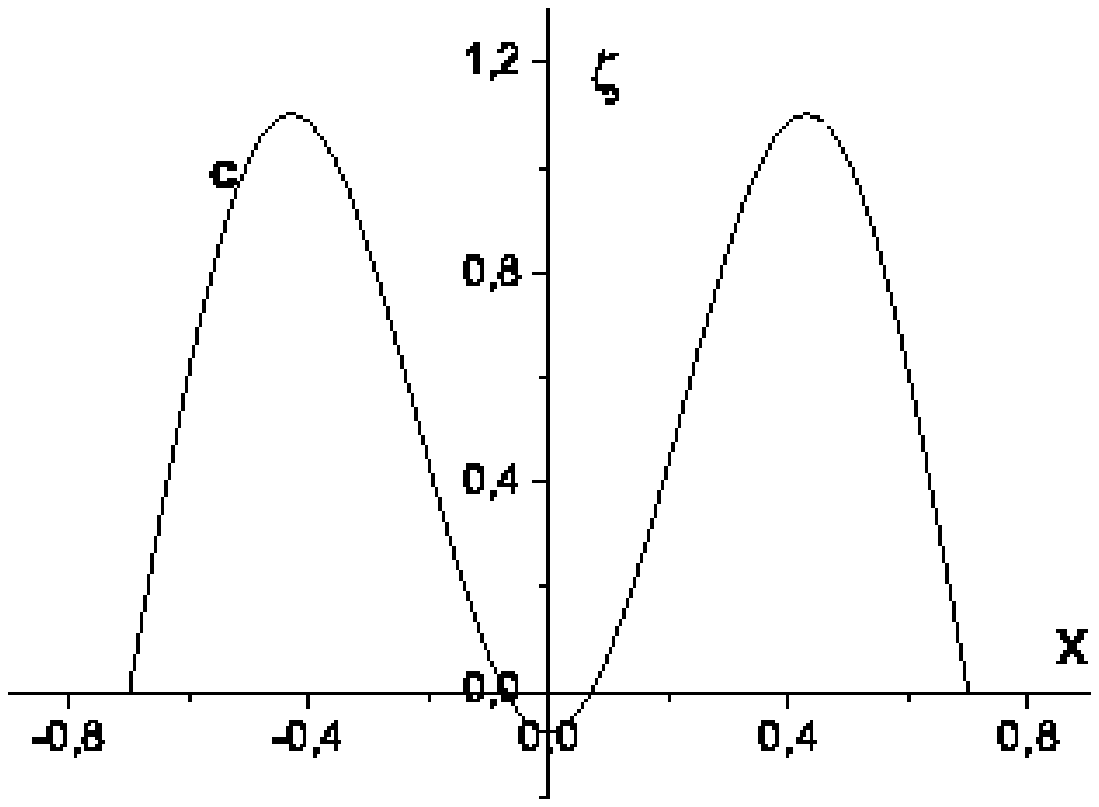}
 \includegraphics[width=4.4cm]{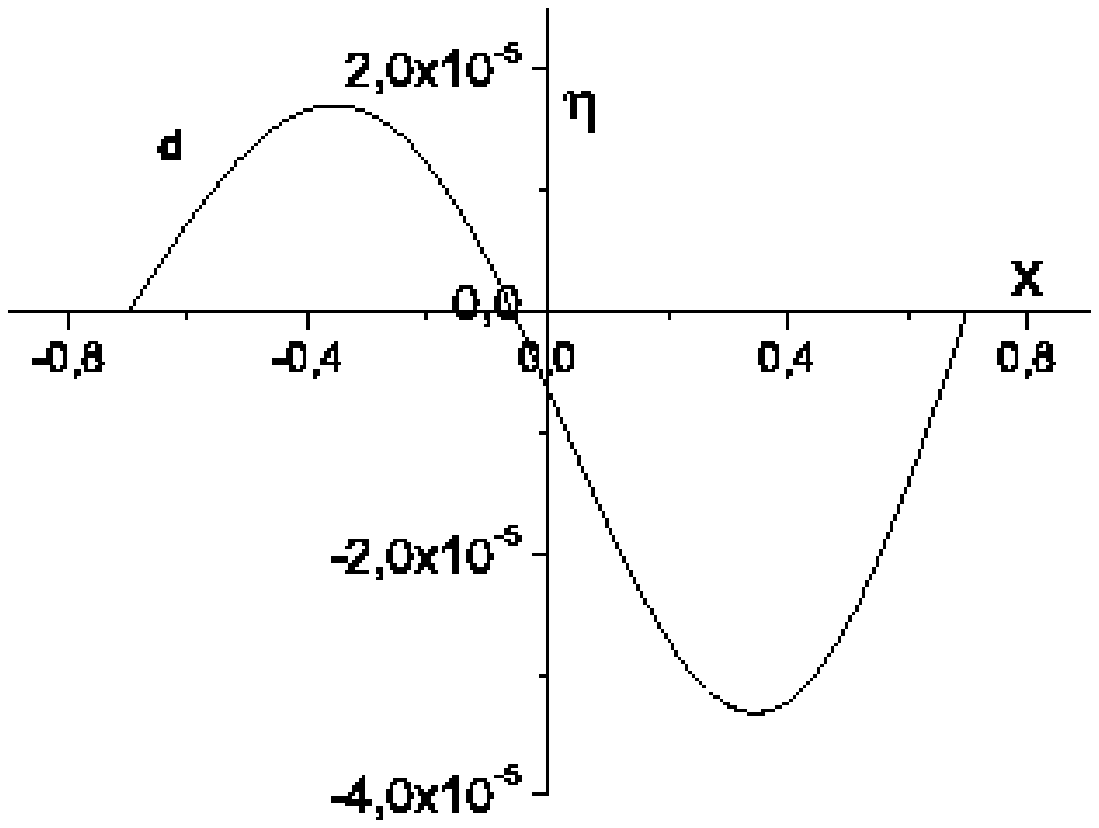}
       \caption{ Mode components corresponding to the first mode
   $P_{1}=36.3 \ min$ for the cases:  a) $\zeta$ and $B_{00}=11G$; b) $\eta$ and $B_{00}=11G$;
    c) $\zeta$ and $B_{00}=100G$; d) $\eta$ and $B_{00}=100G$.}
   \label{fig:dos}
   \end{figure*}

\begin{figure*}[t]
   \includegraphics[width=4.4cm]{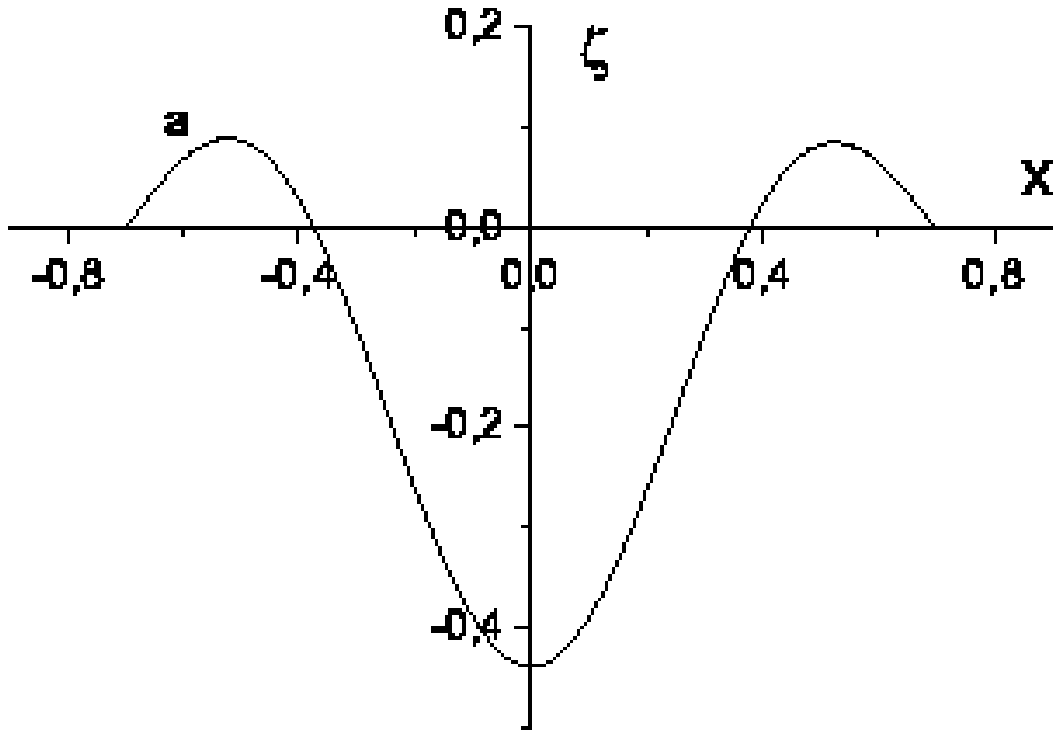}
    \includegraphics[width=4.4cm]{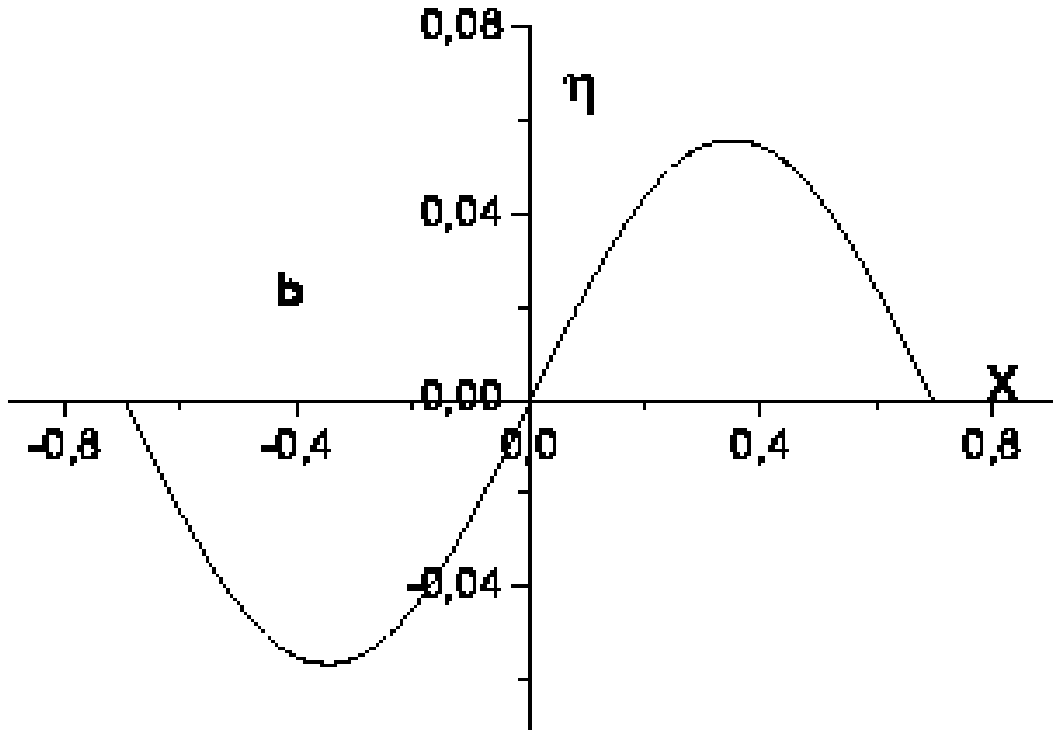}
     \includegraphics[width=4.4cm]{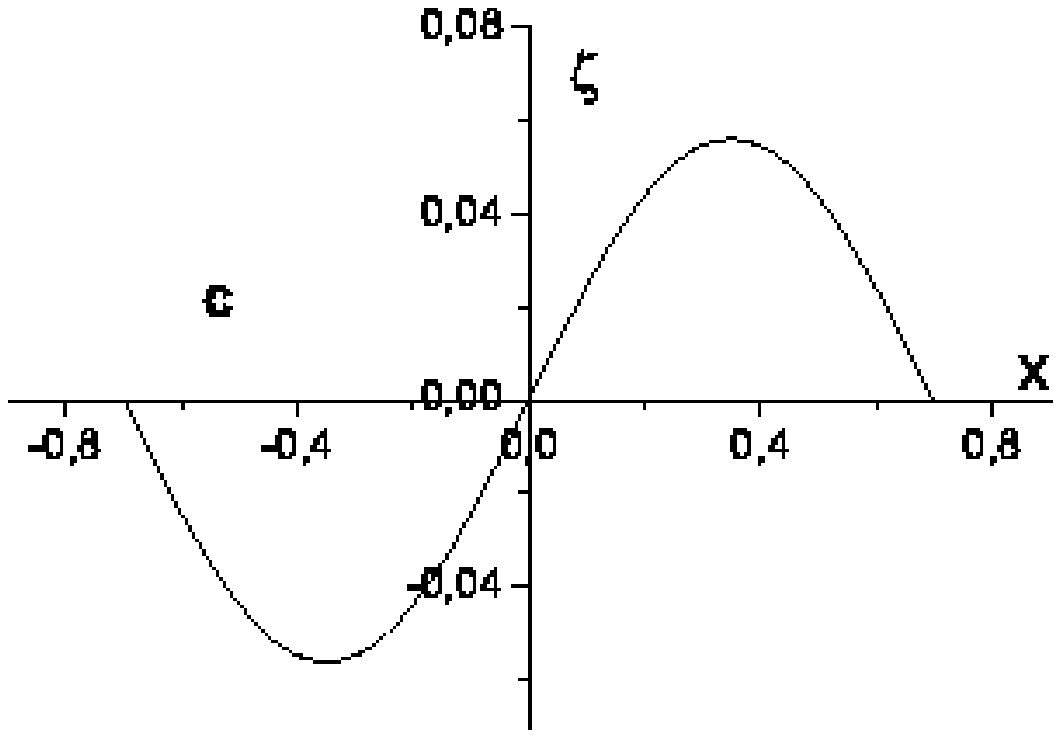}
    \includegraphics[width=4.4cm]{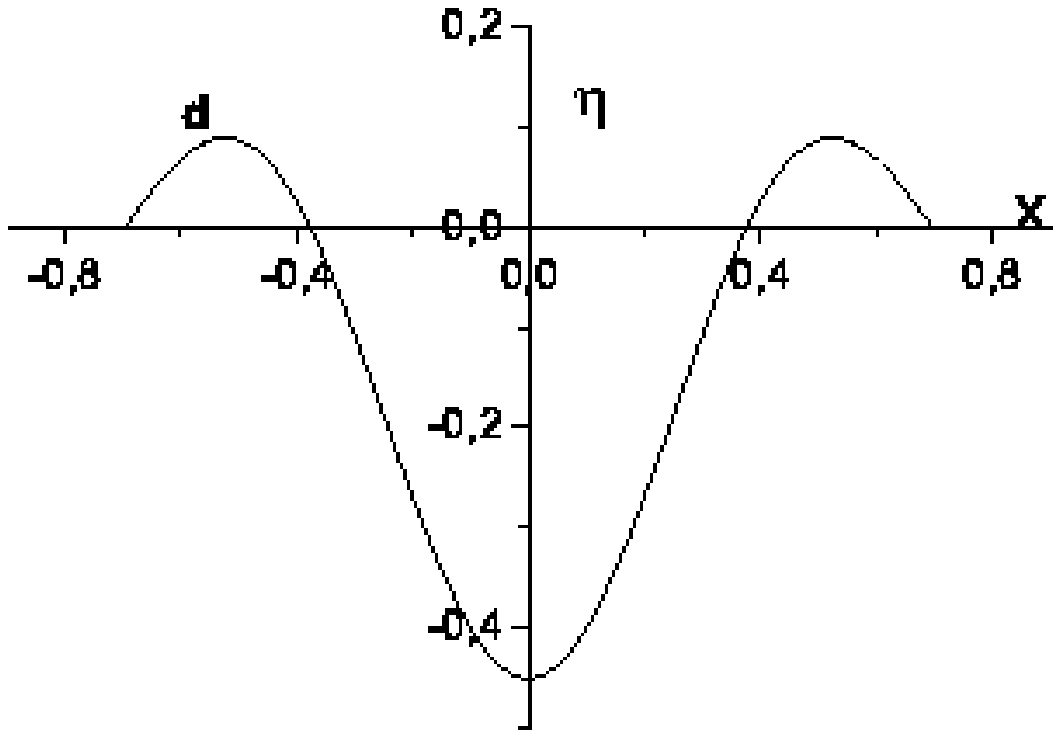}
       \caption{Mode components corresponding to the second
       modes:
   a) $\zeta$ component of $P_{2}=6.6 \ min$ with  $B_{00}=11G$;
    b) $\eta$ component of $P_{2}=6.6 \ min$ with  $B_{00}=11G$;
     c) $\zeta$ component of $P_{2}=0.7 \ min$ with  $B_{00}=11G$;
    d) $\eta$ component of $P_{2}=0.7 \ min$ with  $B_{00}=100G$.}
   \label{fig:tres}
   \end{figure*}
 \begin{figure*}[t]
   \includegraphics[width=4.4cm]{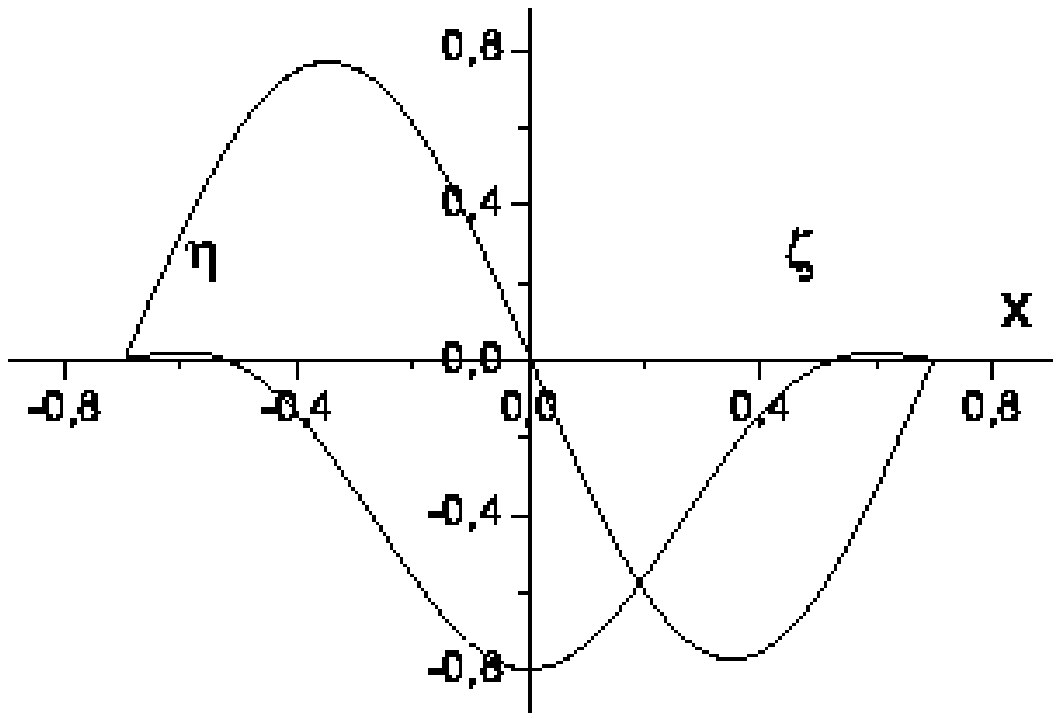}
      \includegraphics[width=4.4cm]{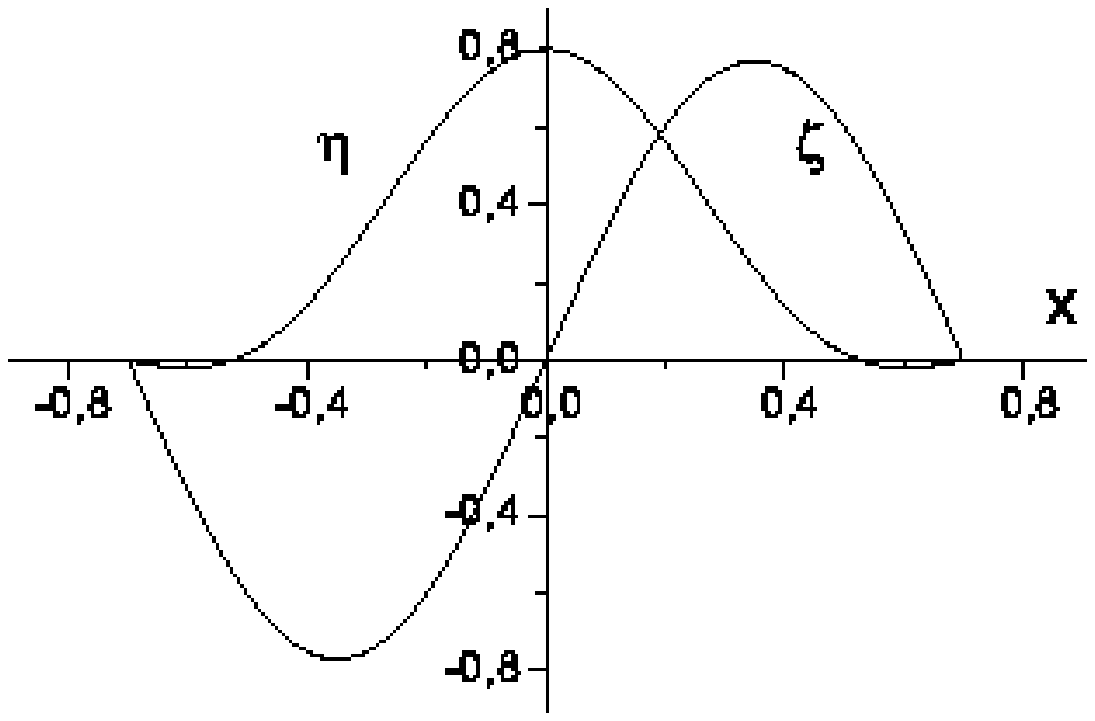}
        \caption{ $\zeta$ and $\eta$ components for the third
        mode.
a) left: $P_{3}=4.3 \ min$ with $B_{00}=11G$ and b) right:
$P_{3}=0.5 \ min$ with $B_{00}=100G$. $\xi_{y}$ has vanishing
values.}
   \label{fig:cuatro}
   \end{figure*}

We analyzed  the structure of the modes with complex values of
$\omega$ as they are  possible candidates for instability
 (Freidberg \citep{fre}). We noted that in the two first
modes the component that is tangent to the magnetic field
$\mid\zeta\mid$ is greater than the component $\mid\eta\mid$ that
is normal to it. This can be seen from Figure~\ref{fig:dos} and
Figure~\ref{fig:tres} where $\zeta$ and $\eta$ are shown for the
cases: $B_{00}=11G$ and $B_{00}=100G$ respectively, also using
$k=0$. The third mode (see Figure~\ref{fig:cuatro}) has comparable
values of $\mid\zeta\mid$ and $\mid\eta\mid$.

\begin{figure}[t]
   \includegraphics[width=5.cm]{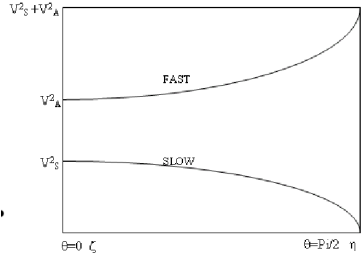}
        \caption{Schematic classification of fast and slow magnetoacoustic waves. $\theta$ is
        the angle between the mode and the magnetic field: $\theta=0$ corresponds to large values
        of $\zeta$ and $\theta=\pi /2$ corresponds to large values of $\eta$.}
   \label{fig:cinco}
   \end{figure}

The fact that for the first mode the two values of $B_{00}$ give
the same time eigenvalue $P_{1}=2\cdot 60\pi / \omega = 36.3 i \
min$ indicates independence from the magnetic structure. This is
consistent with the relative values between the two components in
the two $B_{00}$ cases:
  $\mid\zeta\mid\gg\mid\eta\mid$  (see Figure~\ref{fig:dos}).
 Thus, these magnetoacoustic modes are
more of the acoustic type  $\mid\zeta\mid\gg\mid\eta\mid$ than of
the Alfv\'en  type, i.e. $\mid\zeta\mid\ll\mid\eta\mid$ (see
Figure~\ref{fig:cinco}). Also, the obtained period is included in
a range ($10min<P<60min$) where MHD slow acoustic modes are
expected (Aschwanden \citep{asch}).

Figure~\ref{fig:tres} shows the second  mode for $B_{00}=11G$ and
$B_{00}=100G$ respectively. Also for both cases the
$\mid\zeta\mid$ perturbation is greater than the normal
perturbation  $\mid\eta\mid$ by an order of magnitude.

Figure~\ref{fig:cuatro} show the superposition of
 $\zeta$ and $\eta$ for the third modes corresponding to $P=4.3 \ min$,
 $B_{00}=11G$  and $P=0.5 \ min$, $B_{00}=100G$ respectively. Note
that in these cases, when the component $\eta$ is relevant,
resembling an Alfv\'en  wave, the relation between the eigenvalues
(periods) of the different magnetic fields is $P_{11G}\simeq 10
P_{100G}$, in accordance with the relation between the two values
of $B_{00}(11G)\simeq 10 B_{00}(100G)$ and with the corresponding
values of the Alfv\'en velocities of the medium
$v_{A}=B_{00}/\sqrt{\mu\rho}$.

Figure~\ref{fig:cinco} gives a schematic classification of fast
and slow magnetoacoustic waves from where we can analyze the
behaviour of the modes. The first mode corresponds to $\theta
\approx 0$ and as its eigenvalue is independent of the magnetic
field it gives a slow magnetoacoustic mode. The third mode
corresponds to $0 < \theta < \pi/2$ and as $P_{3,11G}\simeq 10
P_{3,100G}$ it looks like a fast magnetoacoustic mode (Priest
\citep{prie}).

Then, in order to establish the final unstable modes we integrated
 eq.~\ref{17} for each of the normal modes, i.e,
the integrand is the generalized potential energy density.

\begin{figure}[t]
   \includegraphics[width=6.cm]{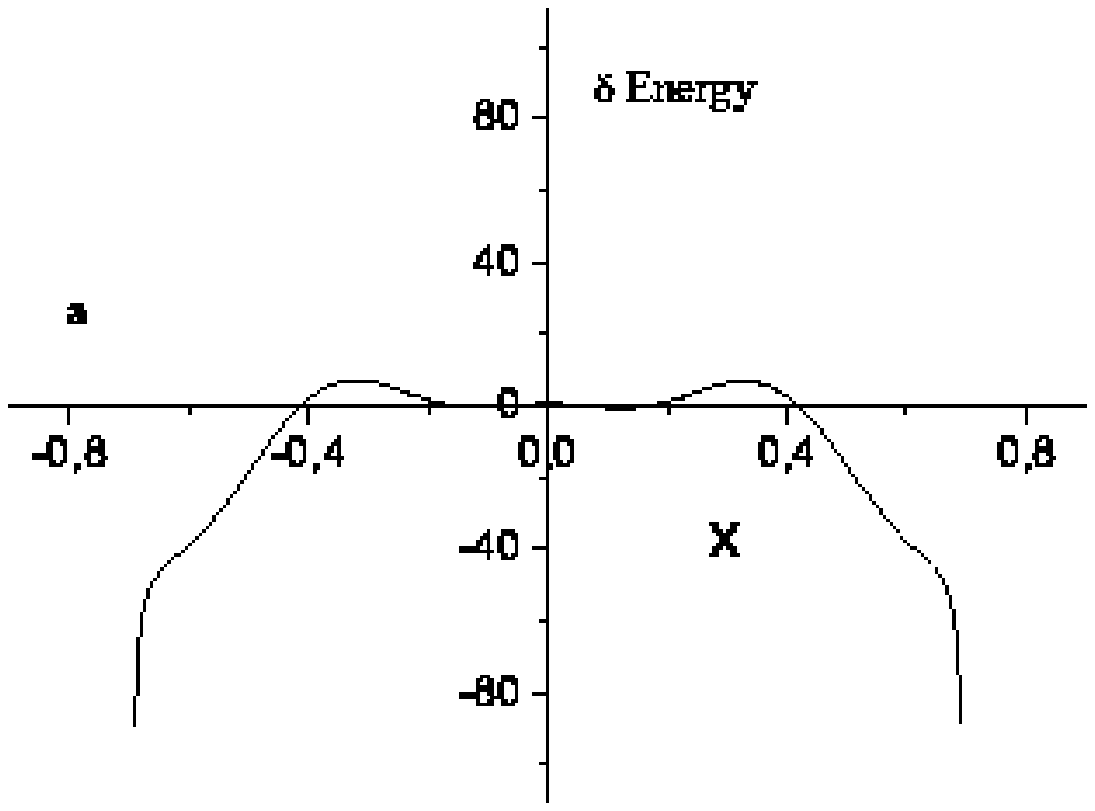}
   \includegraphics[width=6.cm]{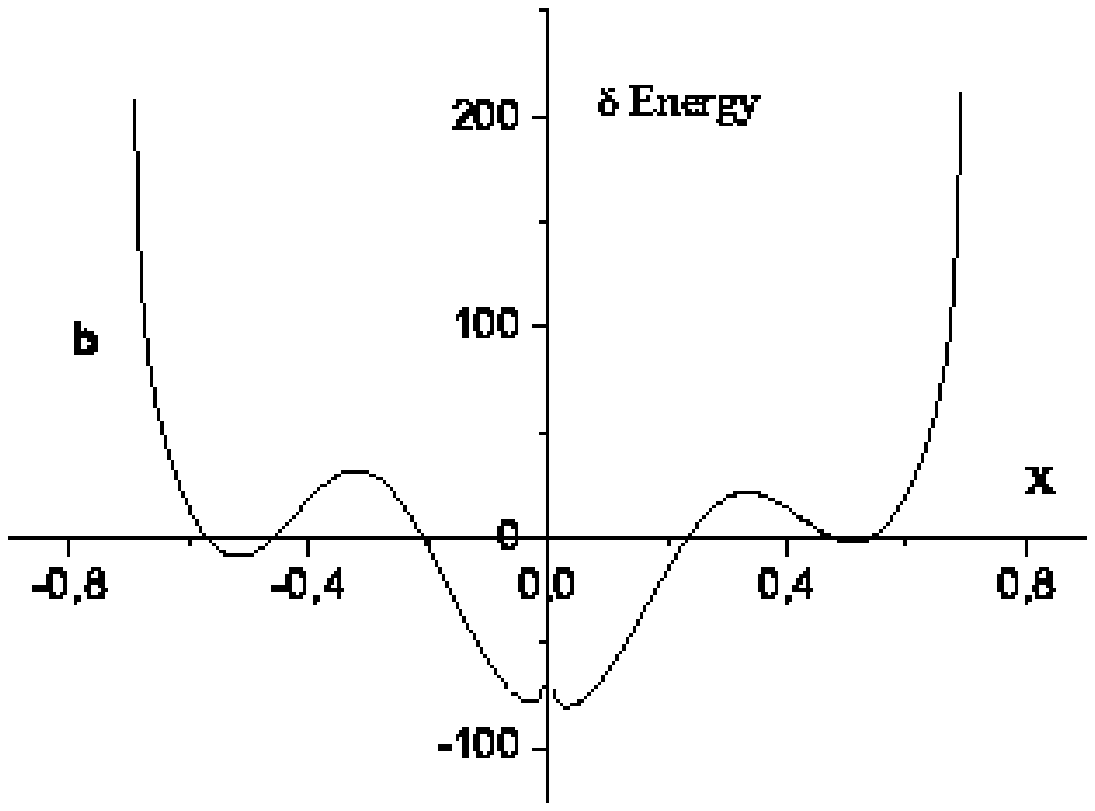}
    \includegraphics[width=6.cm]{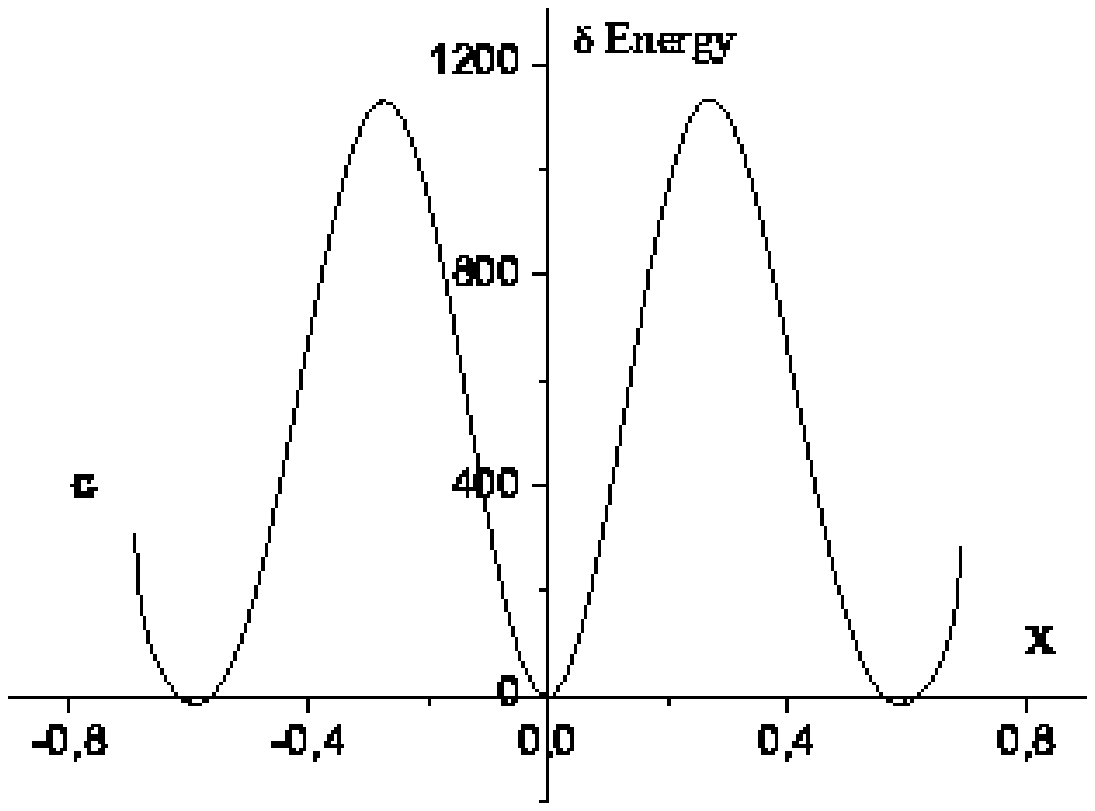}
       \includegraphics[width=6.cm]{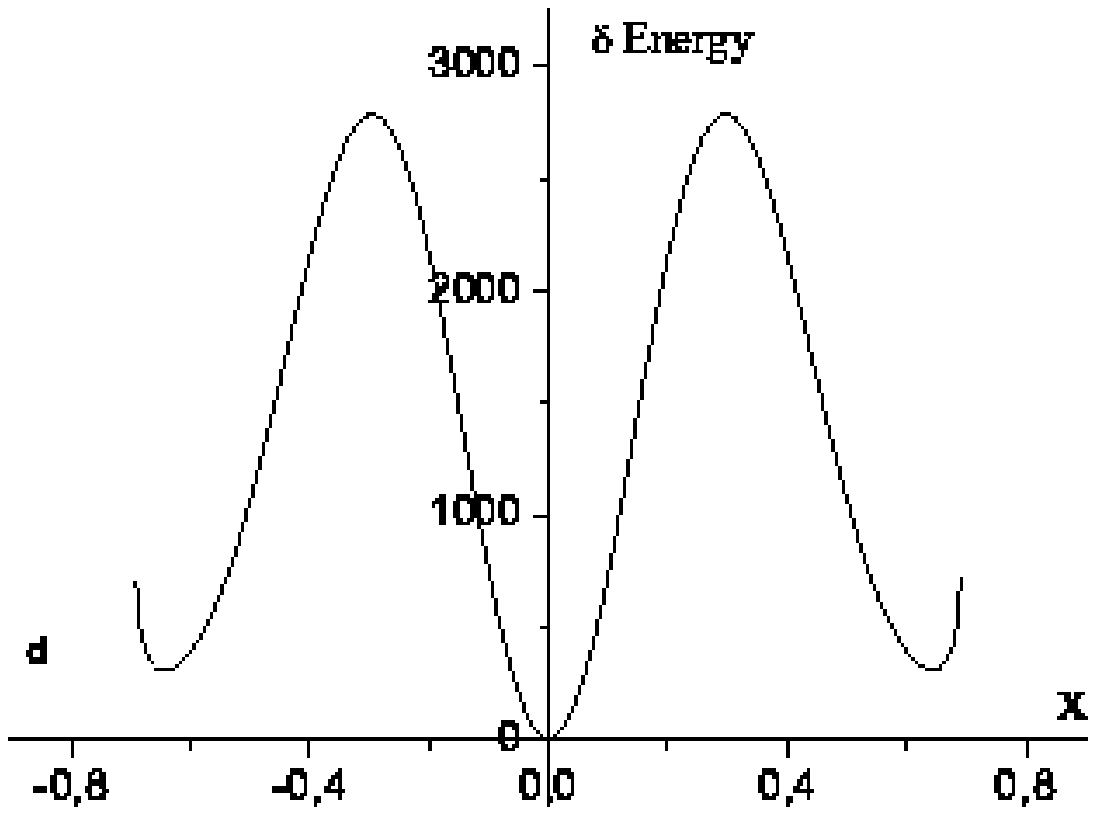}
      \caption{Generalized potential energy density as a function of $x$ for the modes a) $P_{1}=36. \
      min$ with
     $\delta^{2}W_{p}=-16.$; b) $P_{2}=6.6 \ min$ with
     $\delta^{2}W_{p}=2.9$; c) $P_{3}=4.3 \ min$ with  $\delta^{2}W_{p}=6.63$;
     d) $P_{4}=3.4 \ min$ with
     $\delta^{2}W_{p}=2143$,  ($B_{00}=11G$ for all the cases).}
   \label{fig:seis}
   \end{figure}

 Figure~\ref{fig:seis} shows the generalized potential energy density as a
 function of the independent variable $x$ for the three first
 modes (see Table 1), and for the most stable one which was $P_{4}$.
 We show the case  $B_{00}=11G$, the case
with $B_{00}=100G$ has the same functional dependence. Table 3
shows the eigenvalues and the potential energy  for the modes with
complex eigenvalues and for the most stable one.  Note that, even
when $\omega$ has complex values for the three first modes, as
$\delta^{2}W_{p}$ is positive in the second and third case,  the
$P_{1} = 36.3 \ min$ mode is the only unstable one. The fact that,
on the contrary to what happens with the first mode, the other
modes with complex $\omega$ seemed to accumulate at the origin is
an indication of non--real unstable modes.

\begin{figure*}[t]
   \includegraphics[width=4.4cm]{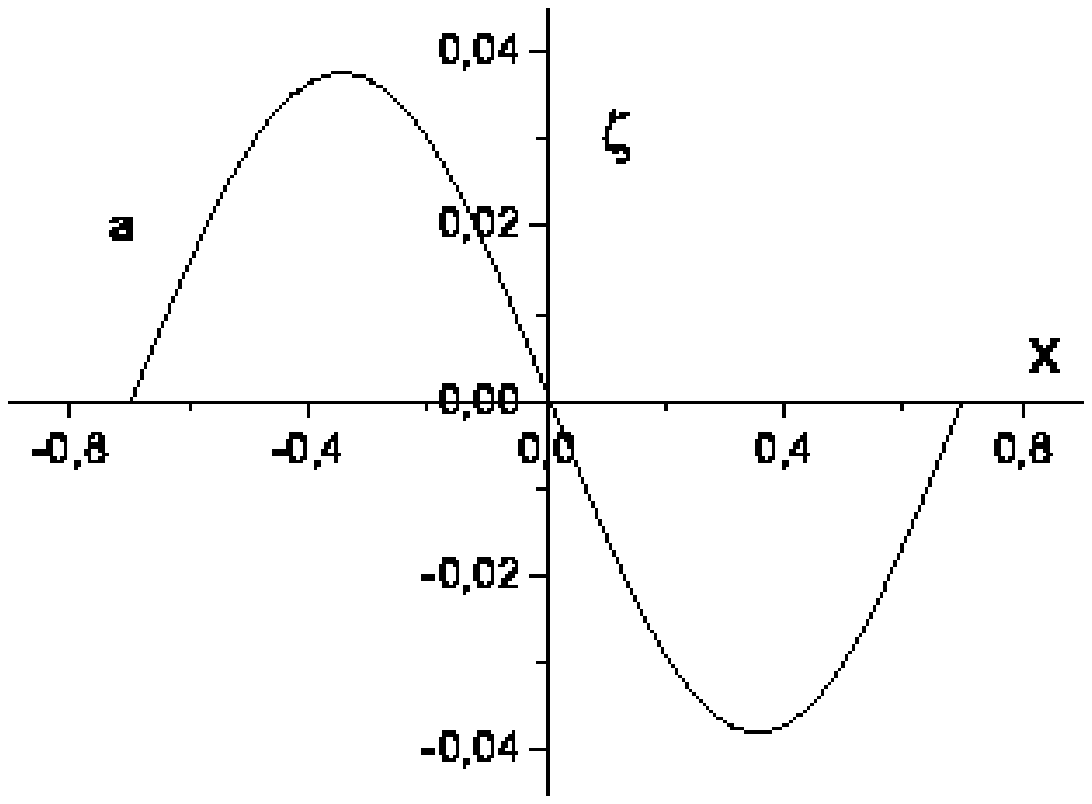}
  \includegraphics[width=4.4cm]{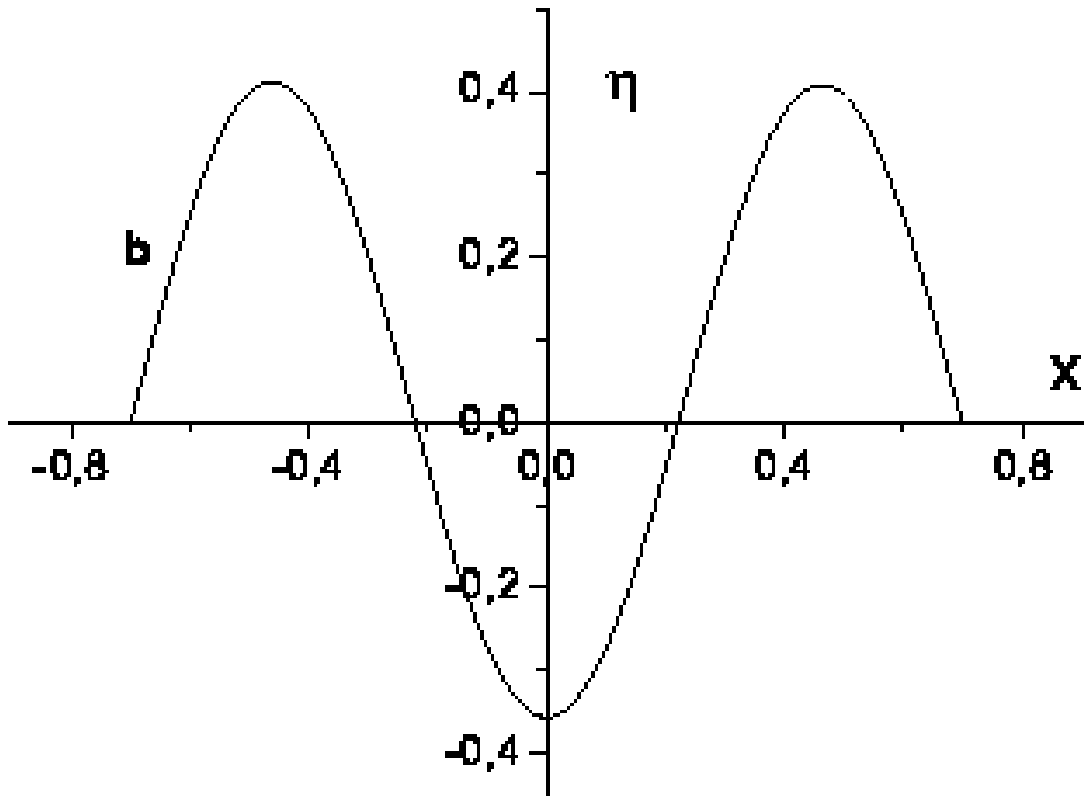}
  \includegraphics[width=4.4cm]{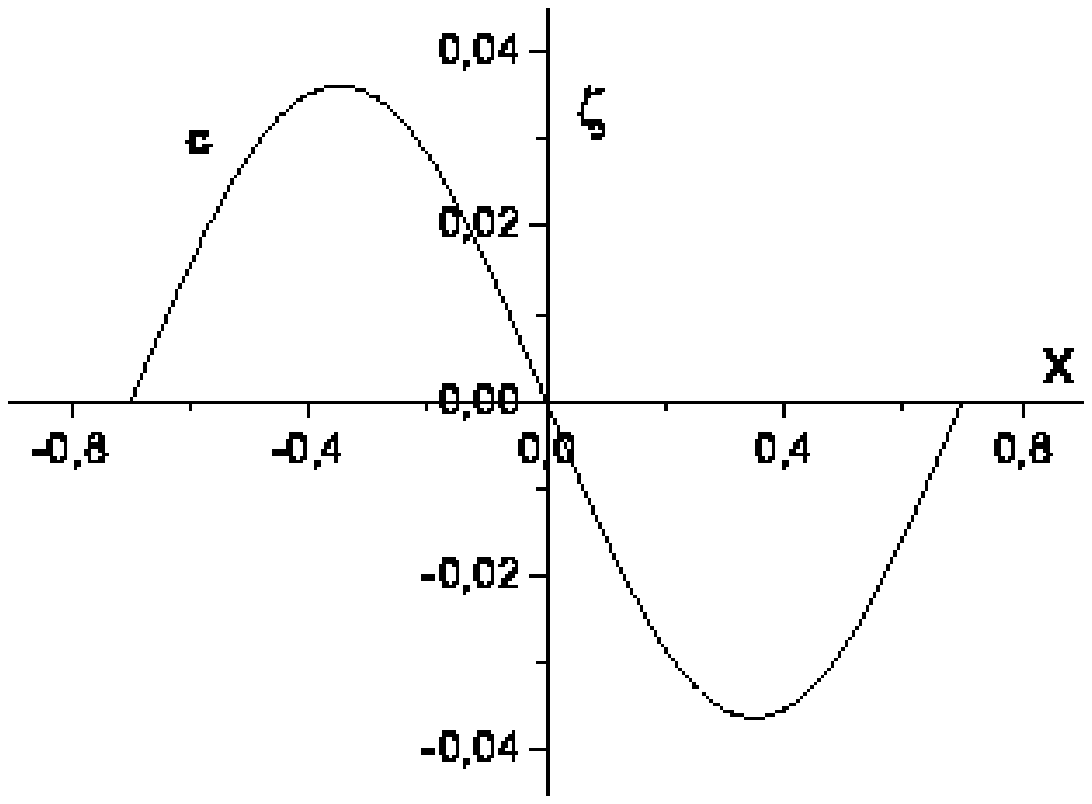}
   \includegraphics[width=4.4cm]{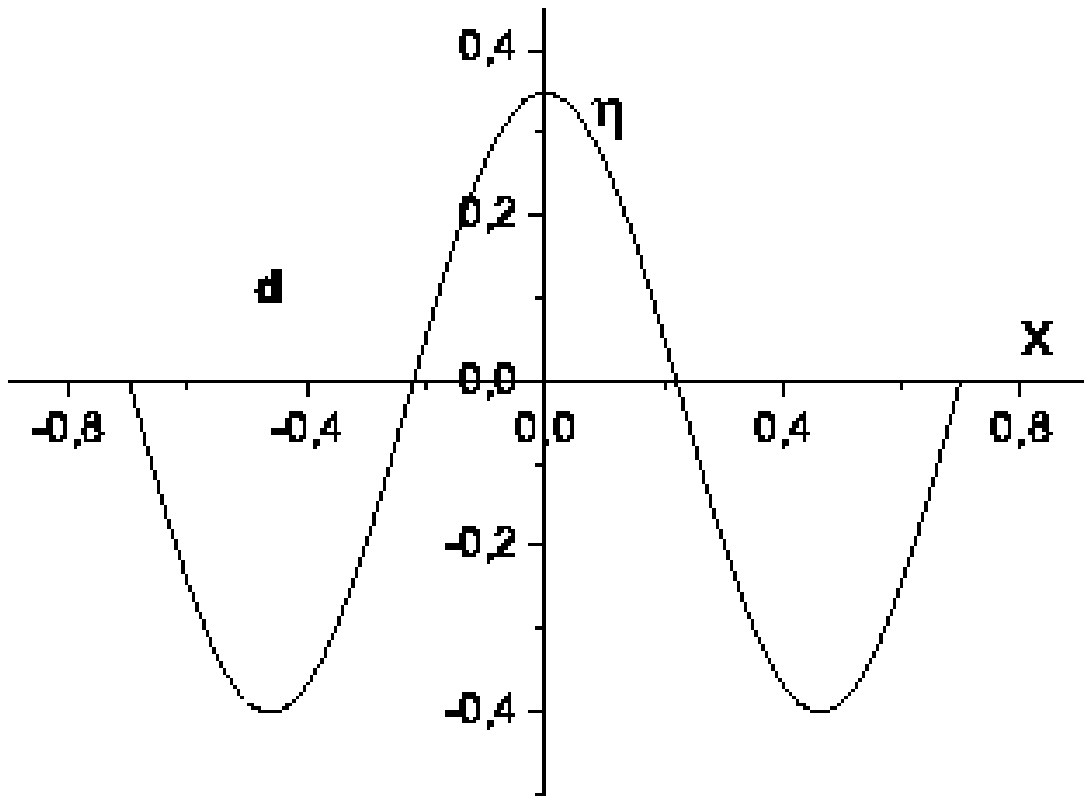}
       \caption{ components of the most stable periods
a) $\zeta$ component of  $P_{4}=3.4 \ min$ with $B_{00}=11G$;  b)
$\eta$ component of  $P_{4}=3.4 \ min$ with $B_{00}=11G$; c)
$\zeta$ component of $P_{4}=0.36 \ min$ with $B_{00}=100G$; d)
$\eta$ component of $P_{4}=0.36 \ min$ with $B_{00}=100G$.}
   \label{fig:siete}
   \end{figure*}

Figure~\ref{fig:siete} shows the structure of the components
$\zeta$ and $\eta$  for the most stable mode $P_{4}$ and for the
two cases: $B_{00}=11G$ and $B_{00}=100G$. Note that
$\mid\eta\mid\geq\mid\zeta\mid$ and that $P_{4,B(11G)}=3. 4 \ min
 \simeq  10 \cdot 0.36 \ min = 10\cdot P_{4,B(100G)}$

 The mode structure of the stable eigenvalues can also be
compared with recent results from the literature. Magnetoacoustic
oscillations of the fast kink type have been studied theoretically
 (Edwin and Roberts  \citep{rob}) and directly observed in EUV wavelengths with TRACE
   (an updated  review of theoretical and observational results in
 Aschwanden  \citep{asch} and references therein).
The observations are usually  modeled by cylinders with a surface
boundary  representing coronal loops. The dispersion relation is
obtained matching the internal and external MHD solutions via the
requirement of continuity of pressure and perpendicular velocity.
As in our model, the observed kink-mode oscillations correspond to
the long-wavelength regime. In coronal conditions the magnetic
field is almost equal inside and outside of the loop and the kink
oscillation speed is almost the  Alfv\'en one depending on the
ratio of external and internal density values, i.e., outside and
inside the loop. On the contrary, our model is performed by
perturbing a magnetic arcade, without considering a cylinder with
different inside and outside conditions. In eleven observational
kink-mode oscillations from which the magnetic field of the events
can be inferred were obtained by Aschwanden et al. \citep{aschaaa}
and \citep{aschbbb}.
 The
comparison of our stable mode data $P_{i>1} $ in the
 $B_{00}=11G$ case is in good agreement with the kink-mode observational results.
The period range (see Table 1), the magnetic strength
($B_{00}=11G$) and the wave speed (Alfv\'en speed) fit the
observations for similar loop densities and loop lengths. Also,
the stable modes $P_{i
>1}$ with $B_{00}=100G$ (see Table 2) have periods that are
comparable with the expected range of fast sausage-mode periods
($P\pm 1-60sec$) and wave speeds of the order of the Alfv\'en
speed (Aschwanden \citep{asch}). However -even when a more
precise comparison requires  a modeling that takes into account
differences between external and internal conditions- it is worth
investigating whether these type of modes could be associated with
more intense magnetic fields in comparison to the associated
kink-mode magnetic fields. This  will be attempted in future work.

A main result regarding stability is that the characteristic time
$\tau= 36 \ min$ in which the instability grows  is large enough
to guarantee a relative permanence of the structure before it
fades away: $\tau \simeq t_{obs}$; where $t_{obs}$ is the typical
characteristic time in which loops seem stable  (see Costa and
Stenborg \citep{cos2}). Thus, even when the non--linear
stationary configuration of Figure~\ref{fig:uno} is unstable it
lasts long enough for the observations to be made. Moreover, we
confirm that the dynamic brightenings usually observed
 could  be due to magnetoacoustic
waves  i.e. the perturbations have short periods in comparison
 with the time that instability occurs:
$ P_{4}= 3.4 \ min$  and $P_{4}=0.36 \ min$ satisfy $P_{4} \ll
\tau$.

Thus, even when further calculation is needed in order to adjust
the characteristic times, it seems that  wave--based models could
be able to describe the scenario of non--isothermal coronal loops
for sufficiently short times comparable with the characteristic
time in which the instability grows and the structure fades away.
A more speculative argument about the relation between wave--based
models and flow--based ones is given in the conclusions.

 \begin{figure}[t]
    \includegraphics[width=5.cm]{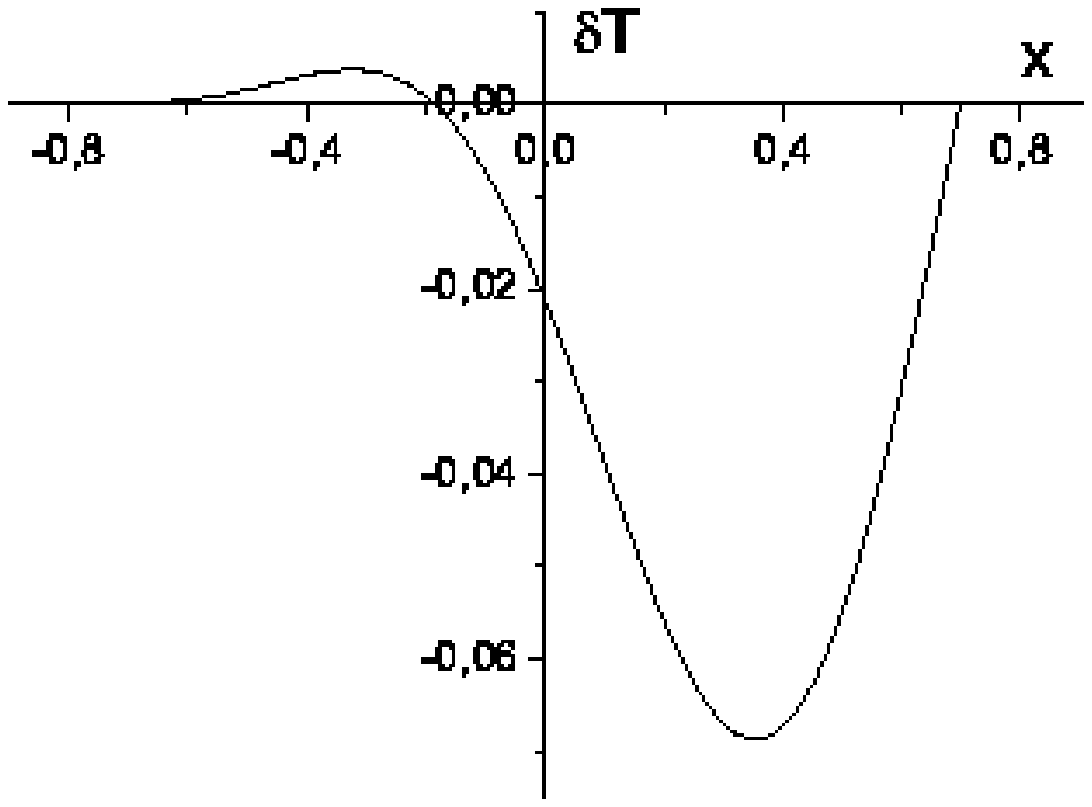}
   \includegraphics[width=5.cm]{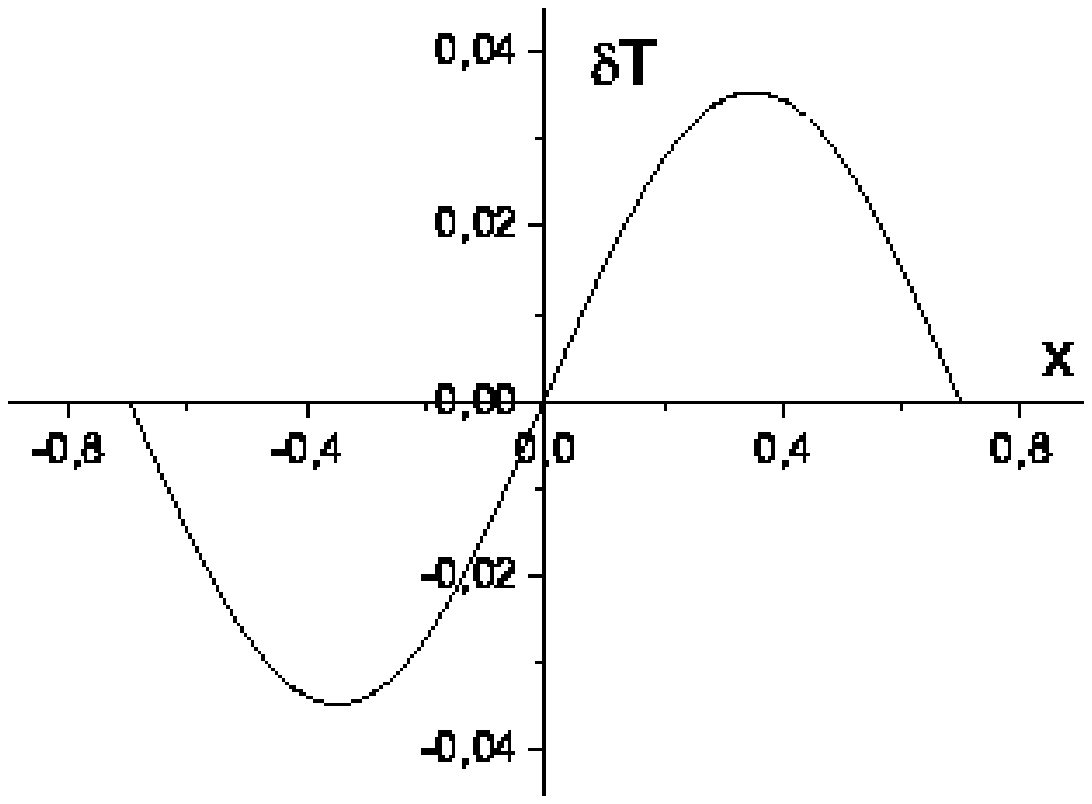}
      \caption{Thermal perturbation ($T_{1}$ component) for the
      cases:
a) $P=36 \ min$;  $B_{00}=11G$ b)  $P=36 \ min$; $B_{00}=100G$
 }
   \label{fig:ocho}
   \end{figure}
    \begin{figure}[t]
   \includegraphics[width=3.5cm]{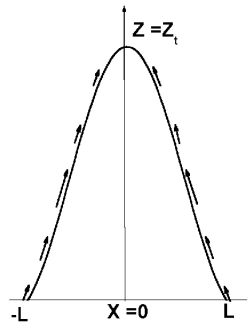}
        \caption{Schematic description of the unstable
        mode superimposed on the magnetic structure. At a definite
        phase the perturbation is always positive, it grows until
        it reaches $x=\pm L/2$, then decreases until it becomes
        zero at $Z=Z_{t}$
 }
   \label{fig:nueve}
   \end{figure}


 \begin{figure}[t]
   \includegraphics[width=5.cm]{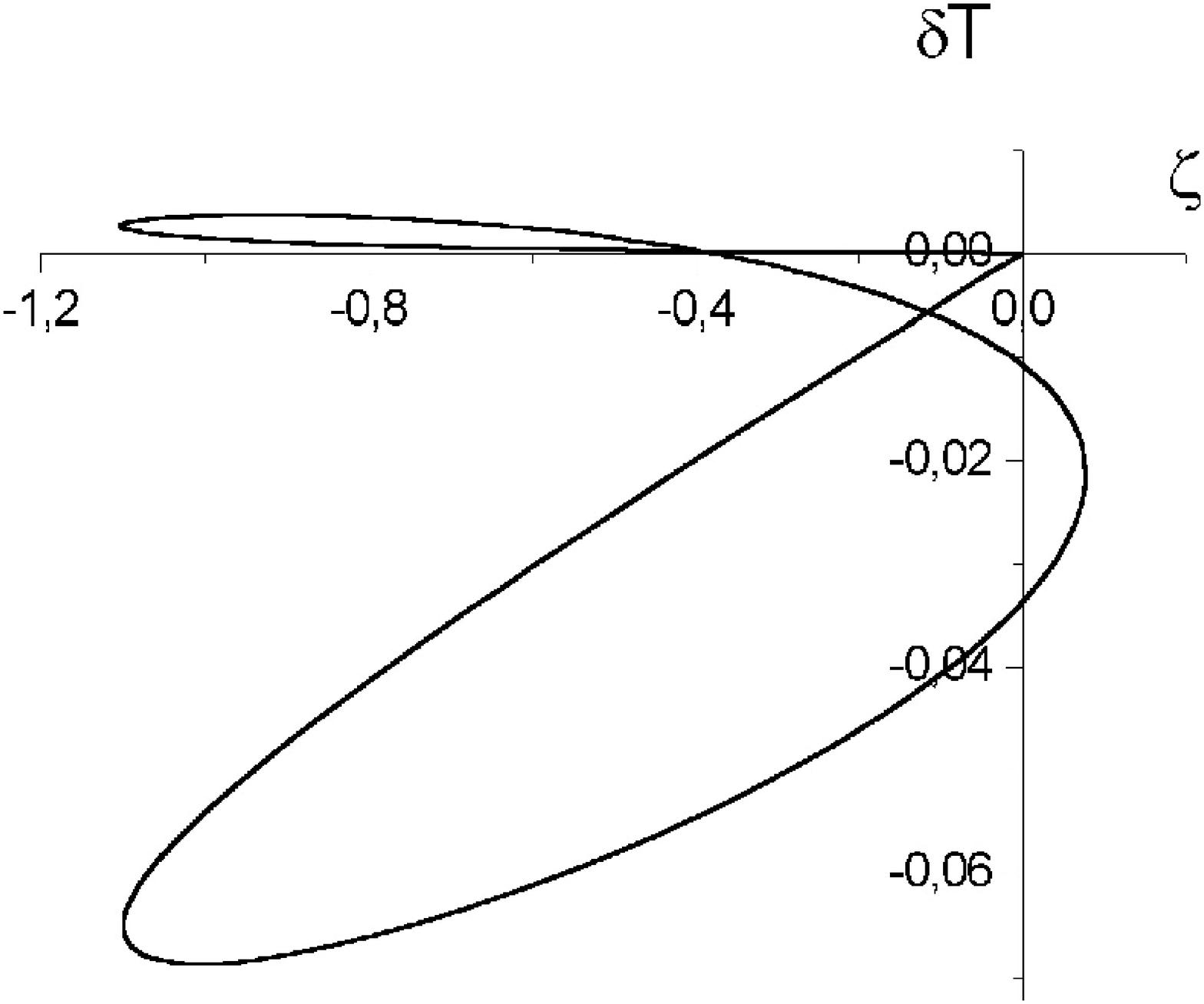}
     \includegraphics[width=5.cm]{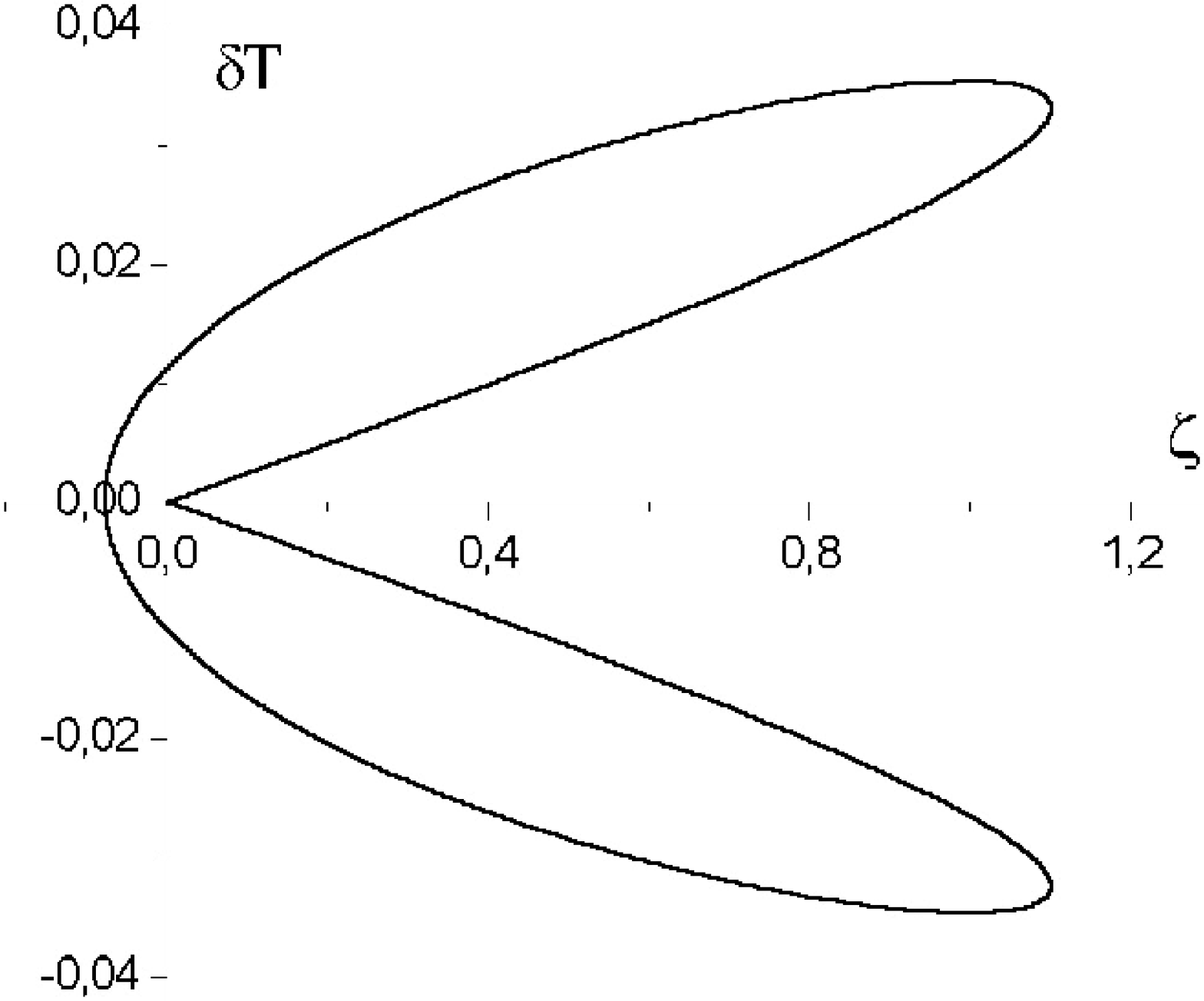}
        \caption{The curve formed by the resulting component perturbations in the vector
        space of perturbations for a) $B_{00}=11G$ and b) $B_{00}=100G$.
 }
   \label{fig:diez}
   \end{figure}

\section{Conclusions}

We investigated -via a thermodynamic energy principle- the
stability of a coronal inhomogeneous loop model in a non--linear
equilibrium state, i.e. a given thermal and magnetic equilibrium
configuration. We also obtained the frequencies and their
associated modes.  The perturbation chosen was of the general type
described by eq.~\ref{31} which allowed the study of a more
complex mode structure with coupled thermal and mechanical
displacements from the equilibrium state. We used a
three--component Fourier basis expansion on the independent
coordinate $x$ to characterize the unstable modes. We obtained
three complex eigenvalues  and six real ones with their
corresponding eigenvectors for each  of the magnetic field values
$B_{00}$  analyzed. The other three modes were discarded. The
definite stability condition of the modes is given by integrating
the generalized potential energy density of eq.~\ref{17}, allowing
the interpretation of long--wavelength disturbances that are
present in inhomogeneous media.

\begin{enumerate}
  \item  We used different values of $k$ ($k=0$, $k=0.5$ and $k=10$)
  to calculate the eigenvectors with complex eigenvalues and for all
  cases
    we obtained
  vanishing values of $\xi_{y}$ with respect to the other perturbed quantities.
  When we repeated the procedure to calculate the modes with real
  eigenvectors
  we obtained small but
  not vanishing values of $\xi_{y}$
   in comparison with the other components. Thus,
   two dimensional loop coronal models with a  temperature gradient of two orders of magnitude  between
  the bottom and top are a good approximation to study the whole
  three dimensional  stability.
    \item  We can classify the structure of the   modes obtained as follows: a) those for which
    $\zeta\gg\eta$ and b) those for which $\eta\geq\zeta$. In the
    first case the  perturbation $v_{1}=\partial\vec{\xi}/\partial
    t$ is almost parallel to the magnetic field and the
    eigenvalue is relatively independent of its intensity resembling the acoustic waves where
    $v_{s}$ is independent of the magnetic field (see Figure~\ref{fig:cinco}). This basic
    longitudinal mode describes an oscillation between parallel
    plasma kinetic energy and plasma internal energy.
    In the
    second case $v_{1}=\partial\vec{\xi}/\partial
    t$ has an important  orthogonal component and the eigenvalue
    varies with the magnetic field
    $P_{11G}\simeq 10 P_{100G}$ resembling the dependence of the
    Alfv\'en waves $v_{A}\simeq B_{00}$. When the wave is nearly transverse it describes an
    oscillation between perpendicular plasma kinetic energy and the combined magnetic
    compressional and line bending energies. Thus,  the
    first case can be thought of as  slow magnetoacoustic waves
    and the second one as fast magnetoacoustic waves. The period
    of the slow magnetoacoustic mode
    is also in accordance with observational data. Between the fast magnetoacoustic
    modes and in the long wavelength regime we distinguish two possible types,
    depending of the strength of the magnetic field.
    For the modes with $P_{i>1}$ and $B_{00}=11G$ we found that the
    period range, the magnetic strength and the speed of the modes
    resemble a fast kink--mode. Also, through the consideration of period
    range values we suggest that modes with $P_{i>1}$ and $B_{00}=100G$
    could be thought of as sausage modes.
  However, to go further with the classification of mode type a
   modeling that takes into consideration differences between the inside and outside
  of  the loop is required. Also,
  the
    non--homogeneous character of the problem  places
    serious limitations on conclusions in relation to stable
    modes.
 \item  We found only one unstable mode with characteristic growing time: $\tau_{u}=36 \ min$. The approximate and
 most stable mode is $P_{4}=3.4 \ min$ for $B_{00}=11G$ and $P_{4}=0.36 \ min$ for
 $B_{00}=100G$. The fact that there is an unstable mode means that the equilibrium
 state is unstable and that wave--based models are not adequate to fit observations.
 However, as $\tau_{u} > P_{4}$  by an order of magnitude or two (depending on the $B_{00}$ value)
 the equilibrium appearance of the loop and the brightening effects of
 the most stable mode could be sustained
 by a characteristic time which is in accordance with
 observational data ($\tau_{u}$ the characteristic time of the instability).
\item  A much more speculative argument, which needs further analysis and  numerical
calculation of the non--linear behaviour of the modes, is as
follows:  the non--linear growth of unstable modes influences the
stable modes (they are called slaves),  the resulting behaviour is
fundamentally governed by the most
 unstable modes. As we obtained a unique unstable mode, of the type of a slow
magnetoacoustic wave, this  indicates  an overall unstable
behaviour governed by the tangential $\zeta$ component
 and the thermal one.
 The thermal component of the unstable
 mode is shown in Figure~\ref{fig:ocho}.
 As a  characteristic wavelength of the components   $\zeta$ and
$T_{1}$ is $L/2$,
 it would be worth investigating whether this
instability could be associated with a limit--cycle solution
generally characterized as a flow--based model (G\'omez et al.
\citep{gom1}; De Groof et al. \citep{deg}). If this is the case,
$\tau$ should be the growth of the instability, before it reaches
its
 non--linear saturated  value, in a  new equilibrium state of an oscillatory type in the $\zeta$ and $T_{1}$
  components. Thus,  both types of models (waves and flow)
   converge in explaining
  the instability of a magnetic structure with long wavelength
  perturbations  of the
  order of the  magnetic structure.
   Also, even when the modes were linearly unstable, the fact that the dominant varying
  components are $T_{1}$ and $\zeta$, with the last one parallel to the magnetic
  field
 could imply that  the magnetic structure (but not the equilibrium state)
 lasts much longer than what is stated by $\tau$. Moreover, this
 is in accordance with the energetic description of the type of perturbation.
 Slow, nearly longitudinal
 magnetoacoustic modes describe a basic  oscillation between
 parallel plasma kinetic energy and plasma internal energy where
 the magnetic energy plays no relevant  role.
 This could
 justify long lasting loop observations with dynamic plasma
 inside.
 Figure~\ref{fig:nueve} is a scheme of the unstable mode superimposed on the magnetic
 structure. In half of the period the  perturbation is always positive and grows until it reaches
 $x=\pm L/2$, then decreases until it becomes zero at
 $x=0$ and $Z=Z_{t}$. The  perturbation gives the tangential velocity
 $v_{1}=\partial\vec{\xi}/\partial
    t$ of the plasma particles at each point of the magnetic configuration.
    Thus, in half of the period, as  described in the figure, the plasma
    is emerges from the chromosphere i.e.,  $\vec{\xi}=\zeta
\mathbf{e}_{n}$. In the other half, the
 perturbation is inverted with respect to the figure,
it  is always negative, i.e. $\vec{\xi}= -\zeta \mathbf{e}_{n}$,
and the plasma particles fall
   into the chromosphere.
  A limit cycle is known to be a closed curve (a cycle) in
  the vector
  space formed by the perturbations. Figure~\ref{fig:diez} shows
  the resulting curve in the  space of  perturbations $(T_{1},
  \zeta)$ for the two magnetic fields studied. It seems that
  only for relatively large values of the magnetic field limit cycle are solutions
   possible.

  \end{enumerate}

\begin{acknowledgements}
This paper is dedicated to the memory of our Professor and guide
Constantino Ferro Font\'an, and also to the memory of our
colleague and friend An\'\i bal Sicardi Schifino.

\end{acknowledgements}

\section{Appendix: Mathematical tools}

The following equations and relations are needed in order to
obtain eq.~\ref{33}

\begin{equation}
 \rho_{t}=\frac{mp}{k_{B}T_{t}}\nonumber \end{equation}
\begin{equation}
\frac{d\rho_{0}}{dx}=\frac{d\rho_{0}}{ds}\frac{ds}{dx}=\triangle\frac{d\rho_{0}}{ds}
\ \  \rightarrow \ \
\frac{dT_{0}}{dx}=\triangle\frac{dT_{0}}{ds}\label{34}
\end{equation}
with $\triangle=\sqrt{1+(z')^{2}}$. From Figure~\ref{fig:uno} it
is easy to show

$$\mathbf{e}_{t}= \frac{\mathbf{e}_{x}}{\triangle}
+\frac{z'}{\triangle}
 \mathbf{e}_{z}; \ \ \mathbf{e}_{n}= -\frac{\mathbf{e}_{x}}{\triangle} +
\frac{z'}{\triangle}
 \mathbf{e}_{z}$$
$$\mathbf{e}_{t}\cdot \mathbf{e}_{x}=\frac{1}{\triangle}; \ \
\mathbf{e}_{t}\cdot \mathbf{e}_{z}=\frac{z'}{\triangle}; \ \
\mathbf{e}_{n}\cdot \mathbf{e}_{x}=\frac{z'}{\triangle}; \ \
\mathbf{e}_{n}\cdot \mathbf{e}_{z}= - \frac{1}{\triangle}$$
 and
$$\mathbf{e}_{t} \times \mathbf{e}_{x}=z'{\triangle}
\mathbf{e}_{y}; \ \ \mathbf{e}_{t} \times
\mathbf{e}_{y}=\mathbf{e}_{n}; \ \ \mathbf{e}_{z}\times
\mathbf{e}_{t}=\frac{\mathbf{e}_{y}}{\triangle}. $$
 Then, the
spatial perturbation  $\xi$ can be written in the Cartesian system
as
\begin{equation} \vec{\xi}=\left[f(\zeta,\eta)
\mathbf{e}_{x}+i \xi_{y}\mathbf{e}_{y}
+g(\zeta,\eta)\mathbf{e}_{z}\right]e^{iky}\label{35}
\end{equation}
taking into account

\begin{equation}
\vec{\xi}\cdot\mathbf{e}_{x}=\left(\frac{\zeta(x)}{\triangle}+\frac{z'}{\triangle}
\eta(x)\right)e^{iky}=f(\zeta,\eta)e^{iky}\label{36}
\end{equation}
and
\begin{equation}
\vec{\xi}\cdot\mathbf{e}_{z}=\left(\frac{\zeta(x)z'}{\triangle}-\frac{1}{\triangle}
\eta(x)\right)e^{iky}=g(\zeta,\eta)e^{iky}.\label{37}
\end{equation}
Then,
\begin{equation}
g(\zeta,\eta)=\eta(x)(1+z'^{2}-f(\zeta,\eta)z')\label{38}
\end{equation}
and
\begin{equation}
D_{x}f=\frac{d}{dx}\left(
\frac{1}{\triangle}\right)\zeta(x)+\frac{1}{\triangle}\frac{d\zeta(x)}{dx}+\frac{d}{dx}\left(\frac{z'}{\triangle}
\eta(x)\frac{z'}{\triangle}\frac{d}{dx}\eta(x)\right)\label{39}
\end{equation}

\bigskip

\bigskip

\noindent $$D_{xx}f=\frac{d^{2}}{dx^{2}}\left(
\frac{1}{\triangle}\right)\zeta(x)+2 \frac{d}{dx}
\frac{1}{\triangle}\frac{d\zeta(x)}{dx}+\frac{1}{\triangle}\frac{d^{2}}{dx^{2}}
\eta(x)+$$
\begin{equation}
\frac{d^{2}}{dx^{2}}\left(\frac{z'}{\triangle}\right)\eta(x)+2\frac{d}{dx}\left(\frac{z'}{\triangle}\right)
\frac{\eta(x)}{dx}+\frac{z'}{\triangle}\frac{d^{2}}{dx^{2}}\eta(x).\label{40}
\end{equation}
\bigskip
\noindent

Finally to obtain a non--dimensional equation the following
changes were made

$$\rho\rightarrow\frac{\rho}{\rho_{t}};  \   \
T\rightarrow\frac{T}{T_{t}}; \   \
\mathbf{B_{0}}\rightarrow\frac{\mathbf{B_{0}}}{B_{00}}; \ \
x,z\rightarrow\frac{x,z}{L}.$$ The other non--dimensional
quantities are obtained immediately from these ones.

 \begin{table}
\begin{tabular}{cccccccccccc}
\hline
 $P_{1}$&$P_{2}$&$P_{3}$&$P_{4}$&$P_{5}$&$P_{6}$&$P_{7}$&$P_{8}$&$P_{9}$&$P_{10}$&$P_{11}$&$P_{12}$
\\
 \hline $36. \ i $&$6.6 \ i$&$4.3 \ i$&$3.4 $&$3.1 $&$1.8$&$1.4 $&$1.3 $&$1.0 $&$0.0$&$0.0$&$0.0$ \\ \hline
\end{tabular}
\caption{\label{tab:table1} Periods associated with the unstable
and stable eigenvalues (minutes) for $B_{00}=11G$.}
\end{table}

 \begin{table}
\begin{tabular}{cccccccccccc}
 $P_{1}$&$
P_{2}$&$P_{3}$&$P_{4}$&$P_{5}$&$P_{6}$&$P_{7}$&$P_{8}$&$P_{9}$&$P_{10}$&$P_{11}$&$P_{12}$
\\ \hline
$36.3 \ i$&$0.7 \ i$&$0.5 \ i$&$0.36$&$0.33 $&$0.2 $&$0.15 $&$0.14
$&$0.1 $&$0.0$&$0.0$&$0.0$\\ \hline
\end{tabular}
\caption{\label{tab:table2} Periods associated with the unstable
and stable eigenvalues (minutes) for $B_{00}=100G$.}
\end{table}
\begin{table}
\begin{tabular}{cccc}
\hline $PotEne_{1}$&$PotEne_{2}$&$PotEner_{3}$&$PotEne_{4}$\\
\hline $-16.9 $&$2.93 $&$663.2 $&$2143.5$\\ \hline
\end{tabular}
\caption{\label{tab:table3} Potential energy (non dimensional) of
the three first modes (complex eigenvalues) and the fourth (most
stable mode) $B_{00}=11G$.}
\end{table}

\end{document}